\documentclass[
 reprint,
superscriptaddress,
 amsmath,amssymb,
 aps,
 prl
]{revtex4-2}

\usepackage{graphicx}
\usepackage{dcolumn}
\usepackage{bm}
\usepackage{xcolor}

\usepackage[T1]{fontenc}
\usepackage[utf8]{inputenc}
\usepackage[english]{babel}
\usepackage{stackengine}
\usepackage[]{graphicx}
\usepackage{ragged2e}
\usepackage[font=small]{caption}
\usepackage{subcaption}
\usepackage{siunitx}

\DeclareCaptionFormat{myformat}{\justifying#1#2#3}
\usepackage{amsmath}
\usepackage{amssymb}
\usepackage{amsfonts}
\usepackage{times}
\usepackage{physics}
\usepackage{siunitx}
\usepackage{bbold}
\usepackage{xcolor}
\usepackage{mathtools}

\usepackage{dsfont}
\usepackage{tabularx}
\usepackage{dutchcal}

\usepackage{wasysym}

\usepackage{hyperref}
\hypersetup{colorlinks=true, citecolor=blue, urlcolor=teal, linkcolor=blue}

\usepackage{tcolorbox}
\tcbset{
    customformula/.style={
        boxrule=0.5pt,
        colback=blue!2!white, 
        colframe=blue!75!black, 
        fonttitle=\bfseries, 
        coltitle=black,
        colbacktitle=blue!5!white, 
        rounded corners=all, 
        boxsep=5pt, 
        left=0pt,
        right=5pt,
        top=5pt,
        bottom=5pt,
        before skip=10pt, 
        after skip=10pt 
    }
}

\definecolor{changered}{rgb}{0.7,0.,0.3}

\allowdisplaybreaks[1]

\usepackage{letltxmacro}
\LetLtxMacro{\ORIGselectlanguage}{\selectlanguage}
\makeatletter
\DeclareRobustCommand{\selectlanguage}[1]{%
  \@ifundefined{alias@\string#1}
    {\ORIGselectlanguage{#1}}
    {\begingroup\edef\x{\endgroup
       \noexpand\ORIGselectlanguage{\@nameuse{alias@#1}}}\x}%
}
\newcommand{\definelanguagealias}[2]{%
  \@namedef{alias@#1}{#2}%
}

\makeatother

\definelanguagealias{en}{english}
\definelanguagealias{EN}{english}
\definelanguagealias{eng}{english}

\begin{document}

\preprint{APS/123-QED}

\title{Probing the nonlocality of Landau levels
in GaAs quantum wells through modified Purcell factors, Lamb shifts and dipole emitted spectra}

\author{Lara Greten} \thanks{Contact author: lara.greten@queensu.ca}
\affiliation{Department of Physics, Engineering Physics and Astronomy,
Queen’s University, Kingston, Canada}
\author{Sabrina Meyer}
\affiliation{Nichtlineare Optik und Quantenelektronik, Institut für Physik und Astronomie, Technische Universität Berlin, Berlin, Germany}
\author{Christina Schröder}
\affiliation{Nichtlineare Optik und Quantenelektronik, Institut für Physik und Astronomie, Technische Universität Berlin, Berlin, Germany}
\author{Andreas Knorr}
\affiliation{Nichtlineare Optik und Quantenelektronik, Institut für Physik und Astronomie, Technische Universität Berlin, Berlin, Germany}
\author{Stephen Hughes}
\affiliation{Department of Physics, Engineering Physics and Astronomy,
Queen’s University, Kingston, Canada}

\date{\today}

\begin{abstract}
In a two-dimensional electron gas, a strong perpendicular magnetic field confines electrons to quantized cyclotron orbits, giving rise to Landau levels with discrete orbit radii.
Even the smallest Landau orbit, set by the magnetic length, spans tens of nanometers for fields of a few Tesla, imposing an intrinsic nonlocal response to electromagnetic excitations.
From a microscopic theory of the nonlocal susceptibility, we derive the Green's function, the central quantity governing all electromagnetic interactions, and evaluate Purcell factors, Lamb shifts, and emission spectra from a proximal dipole emitter beyond the Markov and rotating-wave approximations. Significant nonlocal effects resulting from spatial dispersion of the Landau level response modify the response for experimentally relevant situations up to distances of hundreds of nanometers and, in particular, brighten locally dipole-forbidden transitions due to near-field gradients at multiples of the cyclotron frequency.
The relevant length scales are typical of state-of-the-art nanostructured terahertz architectures, 
and some of our nonlocal features are consistent with 
recent experiments using Landau level polaritons. 
\end{abstract}

\maketitle

\begin{figure}[b]
\centering
    \includegraphics[width=\linewidth]{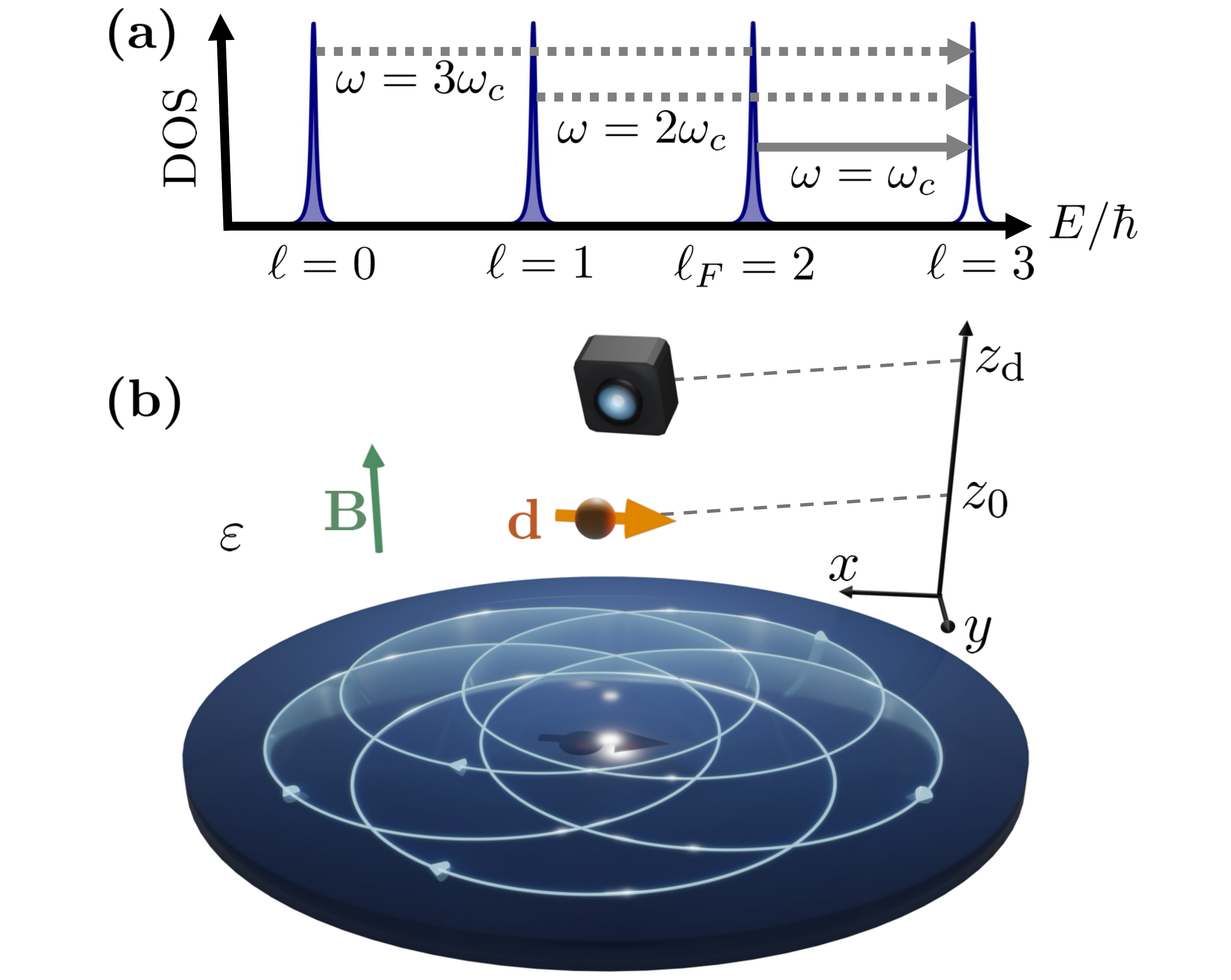}
    \caption{
    \textbf{(a)}
    LL density of states (DOS) for $B=5\,$T, indicating dipole-allowed ($\ell\rightarrow\ell+1$ with $\omega=\omega_c$, solid) and dipole-forbidden ($\omega=2\omega_c$ and $\omega=3\omega_c$, dotted) transitions. Occupied LL up to $\ell_F$ are filled.
    \textbf{(b)} 
    In-plane polarized point dipole $\mathbf{d}$ (orange) at $z_0$ above a Landau-quantized 2DEG (blue) in the $xy$-plane subject to a perpendicular magnetic field $\mathbf{B}$ (green) in a dielectric environment with permittivity $\varepsilon$.
    A detector is indicated at $z_d$.
    }
    \label{fig: intro pic}
\end{figure}

{\it Introduction.}---Landau levels (LLs), which are quantized cyclotron orbits of electrons in a homogeneous, perpendicular magnetic field \cite{landau_diamagnetismus_1930}, rank among the most elementary and experimentally robust macroscopic quantum phenomena \cite{laughlin_quantized_1981,novoselov_two_2005,zhang_experimental_2005}.
In close analogy to harmonic oscillators, they form a ladder of discrete energies $\hbar\omega_\ell=\hbar\omega_c(\ell+1/2)$ with quantum number $\ell$ and cyclotron frequency $\omega_c$, see Fig.~\ref{fig: intro pic}(a).
In particular, THz transitions in two-dimensional electron gases (2DEGs) in high-mobility GaAs quantum wells (QWs) provide a 
clean platform, with large oscillator strengths and long coherence times that, besides the famous quantum Hall effects \cite{tsui_two_1982,palanen_quantized_1982}, have enabled clear demonstrations of ultrastrong light--matter coupling in subwavelength cavities \cite{hagemuller_ultrastrong_2010,scalari_ultrastrong_2012,paravicini-bagliani_magneto-transport_2019,rajabali_ultrastrongly_2022,li_vacuum_2018}. While such cavities are designed to be resonant with the LL transition ($\hbar\omega_c$) in the THz regime, their nanoscale fabrication pushes electric field variations significantly below the micrometer scale, much smaller 
than a typical wavelength.

In a {\it local LL regime}, a spatially homogeneous electric field drives only dipole-allowed cyclotron transitions 
$\ell\rightarrow\ell +1$ at $\omega_c$ 
\cite{wang_direct_2010,palik_infrared_1970}.
However, nonlocal properties introduced by
spatially varying electric fields,
can provide superpositions of
in-plane wavevectors $\mathbf{q}$ that break translational invariance and relax the selection rules.
This leads to:
\textit{(i)} polaritonic nonlocality, where the resonator geometry supplies $\mathbf q$
while the LL response is still treated {\it locally}
\cite{Rajabali_polaritonic_2021,paravicini-bagliani_magneto-transport_2019},  and \textit{(ii)} access to electronic \textit{LL nonlocality}~\cite{endo_cavitymediated_2025,andolina_quantum_2026}---an intrinsic, single-particle effect based on the finite extent of cyclotron orbits, i.e., the Larmor radius $r_c=l_B\sqrt{2\ell+1}$ set by the magnetic length $l_B=\sqrt{\hbar/(eB)}$.
Here, the response cannot be written as $\mathbf{P}(\mathbf{r})=\varepsilon_0\boldsymbol{\chi}(\mathbf{r})\cdot\mathbf{E}(\mathbf{r})$. Instead, the electronic spatial dispersion requires $\mathbf{P}(\mathbf{r})=\varepsilon_0\int\boldsymbol{\chi}(\mathbf{r}-\mathbf{r}')\cdot\mathbf{E}(\mathbf{r}')$.
This {\it LL nonlocality} should also be distinguished from plasmonic nonlocality that, in visible-range plasmonic nanostructures, becomes relevant only on few-nanometer scales (or less)~\cite{mortensen_generalized_2014,raza_nonlocal_2015,greten_strong_2024,eriksen_nonlocal_2024,ciraci_probing_2012}.

Certain nonlocal LL effects have been known for decades \cite{chiu_plasma_1974}, in contexts such as acoustic waves~\cite{greene_linear_1969,kukushkin_dispersion_2009}, scattering of extreme-ultraviolet light~\cite{meyer_proposal_2026}, disorder and impurities~\cite{pinczuk_observation_1988}, Coulomb interactions~\cite{kallin_excitations_1984,kallin_many-body_1985,maag_coherent_2016} and Hall viscosity~\cite{hoyos_hall_2012}.
For graphene, LL nonlocality was shown to modify the Purcell factor at few-nanometer distances \cite{ma_local_2022,eriksen_chiral_2025}, while recent theory suggests that locally forbidden $n\omega_c$ transitions contribute to ultrastrong coupling of graphene in 
plasmonic cavities~\cite{andolina_quantum_2026}.

Signatures of LL nonlocality have already been 
observed~\cite{Rajabali_polaritonic_2021},
using near-field apertures with GaAs QWs, where,
depending on the length scale of the nanopatterning, additional spectral features appear near $1.6\,\omega_c$; however, the theory 
applied a local dielectric LL model~\cite{Rajabali_polaritonic_2021}, and thus a complementary microscopic description is still lacking to place such observations on a firm footing.

To quantify the impact of LL nonlocality on electromagnetic  field observables,
we derive a microscopic, fully retarded Green's function (GF) including a Landau-quantized 2DEG, beyond rotating-wave and Markov approximations.
The electromagnetic GF 
is a powerful tool in quantum field theory~\cite{dung_three-dimensional_1998,joulain_surface_2005,vogel_quantum_2006,Yao_ultrahigh_2009}: it underlies the fluctuation-dissipation theorem and governs a wide range of electromagnetic and quantum-optical phenomena, including dipole-dipole interactions, Casimir/van der Waals forces~\cite{cysne_tuning_2014,muniz_casimir_2021}, and radiative heat transfer~\cite{wu_active_2019}.

We then use the GF to evaluate Purcell factors~\cite{todorov_purcell_2007,wu_strong_2019,rathje_coupling_2023,li_purcell_2026}, Lamb shifts, and emitted spectra~\cite{van_vlack_spontaneous_2012} of a proximal quantum emitter, see Fig.~\ref{fig: intro pic}(b).
We demonstrate a clear breakdown of local LL theory on these obervables, and introduce several new nonlocal effects
that should be observables with present day  experiments, some of which have already been noticed ~\cite{Rajabali_polaritonic_2021}, but without supporting theory:
Our self-consistent theory supports a collective magnetoplasmon \cite{RPA,richards_inelastic_2000,yahniuk_strongly_2026} resonance, which has no counterpart in a local material model.
Moreover, we find that an evanescent near-field \textit{brightens locally forbidden transitions at multiples of $\omega_c$}, see Fig.~\ref{fig: intro pic}(a). 
We stress that such features are entirely linear, and thus distinct from 
nonlinear photovoltage signals at $2\omega_c$~\cite{volkov_bernstein_2014,bialek_photoresponse_2015,bandurin_cyclotron_2022}.

Our significant findings are:
(i)
nonlocal Purcell factors that exceed the local predictions by up to two orders of magnitude at $2\omega_c$, 
for separations out to several hundred nanometers; (ii) the environment-induced Lamb shift can 
change sign with nonlocal effects;
(iii) the dipole-allowed $\omega_c$ transition acquires sizeable nonlocal corrections for distances up to $\approx100\,$nm for typical experimental realizations. 
Our results show that a local material model becomes inadequate if the electric field varies over the spatial extent of the occupied cyclotron orbits
which is typical with state-of-the-art nanostructured THz architectures~\cite{Rajabali_polaritonic_2021,kim_terahertz_2023,andberger_terahertz_2024,timmer_ultrafast_2026} and far beyond common plasmonic nonlocality length scales. This is especially relevant to the ultrastrong-coupling regime, where off-resonant transitions cannot be neglected~\cite{ciuti_quantum_2005,frisk_kockum_ultrastrong_2019,forn-diaz_ultrastrong_2019}, but a local LL response has commonly been presumed sufficient.

\begin{figure*}[t] 
\centering
    \includegraphics[width=1\linewidth]{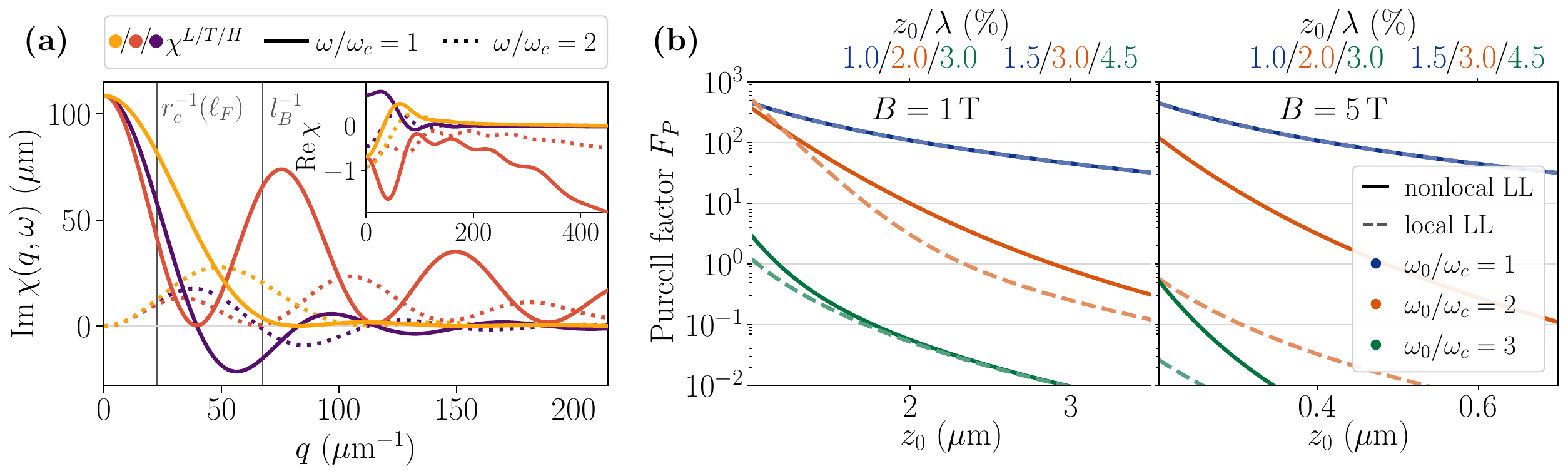}
    \caption{
    \textbf{
    (a)} Imaginary (inset: real) part of the longitudinal ($L$), transverse ($T$) and Hall ($H$) nonlocal susceptibility $\chi^{L,T,H}(q;\omega)$ of a Landau-quantized 2DEG (GaAs quantum well) for $B=3\,$T and temperature $T = 10\,$K over in-plane wavevector $q$ for $\omega=\omega_c$ (solid) and $\omega=2\omega_c$ (dashed).
    Grey vertical lines indicate the inverse maximal, $r_c(\ell_F)$, and minimal, $r_c(\ell=0)=l_B$, Larmor radius.
    \textbf{(b)} Purcell factor for a parallel dipole with frequency $\omega_0$ at $z_0$ above a Landau-quantized 2DEG for $B=1\,$T (left) and $B=5\,$T (right). The dipole is chosen to be resonant with the dipole allowed transition $\omega_0 = \omega_c(B)$ (solid blue), and two dipole forbidden transitions, $\omega_0 = 2\omega_c(B)$ (solid orange) and $\omega_0 = 3\omega_c(B)$ (solid green).
     For each $B$,  $z_0$ constitutes a few percent of the wavelength $\lambda = {c_0}/{(\sqrt{\varepsilon}\omega_0)}$, in a medium with $\varepsilon=12.9$.
     Results based on the local LL response are shown with a dashed line. For the shown $z_0$ and $\omega_0 = \omega_c(B)$, the local approximation coincides with the nonlocal response. {\it Significant deviations up to two orders of magnitude occur for dipole forbidden transitions.}} 
    \label{fig: qs Greens imag}
\end{figure*}

{\it Theory.}---The Hamiltonian for the non-interacting 2DEG is
\begin{align}
    H_0 = \sum_{j,j^\prime} \langle j \vert  \frac{(\mathbf{p}+e\mathbf{A}(\mathbf{r},t))^2}{2M}
    \vert j^\prime \rangle
    c_j^\dagger c_{j^\prime} + V_\text{con} (z),
    \label{Ham:UsualHam}
\end{align}
with
annihilation (creation) operator $c_j^{(\dagger)}$,
effective mass $M$ \cite{haken_quantenfeldtheorie_1993}, elementary charge $e>0$,
and QW confinement potential $V_\text{con} (z)$.
The vector potential,
 $   \mathbf{A}(\mathbf{r},t) = \mathbf{A}_B(\mathbf{r}) + \mathbf{A}_0(\mathbf{r},t)$,
comprises a homogeneous, perpendicular magnetic field $\mathbf{B} = B\mathbf{e}_z$ in the symmetric gauge \cite{prange_quantum_1990,girvin_modern_2019,ciftja_detailed_2020}, $\mathbf{A}_B(\mathbf{r}) = \frac{1}{2} \mathbf{B} \times \mathbf{r}$, while
$\mathbf{E}(\mathbf{r},t)= - \dot{\mathbf{A}}_0(\mathbf{r},t)$
is in the temporal gauge \cite{greene_linear_1969}.
The wave functions,
$\psi_{j=\{\ell,m\}}(\textbf{r}) = \langle \mathbf{r}\vert j \rangle =\varphi_j(\boldsymbol{\rho} )\xi(z)$
with
$\mathbf{r} = (\boldsymbol{\rho} ,  z )^\mathsf{T}$,
solve the out-of-plane $(z)$ confinement potential $V_\text{con}$ via $\xi$ and account for the in-plane $(\boldsymbol{\rho})$ LL by $\varphi_{j=\{ \ell,m\} }$ with the positive integer quantum numbers for energy ($\ell$) and spatial degeneracy $(m)$.
Strongly correlated phases due to Coulomb interactions \cite{haldane_nobel_2017,macdonald_magnetoplasmon_1985}, such 
as Wigner crystallization and the fractional quantum Hall effect \cite{girvin_magneto_1986,murthy_hamiltonian_2003,papic_fractional_2022,jain_composite_2015,pan_experimental_2008}, and the spin-degree of freedom leading to skyrmion excitations \cite{sondhi_skyrmions_1993,balram_fractionally_2015}, typically observed in transport measurements or Raman spectroscopy, are excluded from our description.

For a real vector potential $\mathbf{A}\in\mathbb{R}^3$, the current density:
\begin{align}\label{eq:current_density_final_Pi}
    \mathbf j(\mathbf r,t)
   & =
    -\frac{e}{2M}
    \sum_{j,j'}
    \Big[
        \psi_j^*(\mathbf r)\,
        \bigl(\mathbf p
        +e\mathbf A(\mathbf r,t)\bigr)\psi_{j'}(\mathbf r)\\
        &+
        \psi_{j'}(\mathbf r)\,
        \Bigl(\bigl(\mathbf p
        +e\mathbf A(\mathbf r,t)\bigr)\psi_j(\mathbf r)\Bigr)^*\,
    \Big]
    (c_j^\dagger c_{j'})(t),
    \nonumber
\end{align}
follows in the symmetrized form,
since \(\mathbf p\) and \(\mathbf A_0\) do not necessarily commute.
We average the microscopic structure of the GaAs unit cell within which variations of Landau wave functions \(\varphi_j(\boldsymbol{\rho})\), the confinement envelope \(\xi(z)\), and the electric field $\mathbf{E}(\mathbf{r},t)$ are assumed to be negligible \cite{harper_finite-wave-vector_2018}.
In linear response, we evaluate the transition amplitudes \(\langle c_j^\dagger c_{j'}\rangle\) to first order in \(\mathbf A_0\). The equilibrium LL occupations follow a Fermi--Dirac distribution with a $B$-dependent chemical potential \cite{vagner_ideally_1983}. We introduce a phenomenological damping rate \(\gamma\)
accounting for finite electron mobility and coherence times \cite{curtis_cyclotron_2016,wang_direct_2010}.
Finally, we define the surface current density
$
\mathbf{j}^{\text{2D}}(\boldsymbol{\rho},t)
=
\mathbf{j}(\mathbf{r},t)/|\xi(z)|^2
$.
Since the magnetic field is spatially constant, this yields a two-dimensional nonlocal LL susceptibility \(\boldsymbol{\chi}(\boldsymbol{\rho},\boldsymbol{\rho}';\omega)= \boldsymbol{\chi}(\boldsymbol{\rho}-\boldsymbol{\rho}';\omega)\), which is translationally invariant for an infinite sample.

It is important to highlight that
despite this translational invariance,
LL can still allow for nonlocal behavior. 
For example, a dipole response
at ${\bf r}$, is affected by
fields over ${\bf r}'$ in the vicinity, such that even the photonic LDOS 
can have nonlocal LL interactions. Equivalently, this
nonlocality can be viewed as a feature of
spatial dispersion, mathematically captured 
by 
$\boldsymbol{\chi}(\mathbf q\neq0,\omega)$
with an in-plane wavevector $\mathbf{q}$.

We identify the susceptibility connecting the surface dipole density $\mathbf{P}^\text{2D}=\partial_t\mathbf{j}^{2\mathrm D}$ with the electric field, 
\begin{equation}
    \mathbf{P}^{2\mathrm D}(\mathbf{q};\omega)
    =
    \varepsilon_0
    \boldsymbol{\chi}(\mathbf{q};\omega)
    \cdot
    \mathbf{E}(\mathbf{q},z=0;\omega).
    \label{eq: definition of chi}
\end{equation}
Rotating the in-plane coordinate system along $\mathbf{q}$, longitudinal $(L)$, transverse $(T)$ and Hall $(H)$ contributions appear as entries of the nonlocal susceptibility:
\begin{equation}
\label{eq: LL nonlocal susceptibility tensor wavevector space}
\boldsymbol{\chi}(\mathbf{q},\omega)
=
\begin{pmatrix}
\chi^L({q},\omega)
& 
-i\chi^H({q},\omega)\\
i\chi^H({q},\omega)
&
\chi^T({q},\omega)
\end{pmatrix},
\end{equation}
which depend only on the absolute value $q=|\mathbf{q}|$.
Explicit expressions (full derivations)
are shown in the End Matter (Supplemental Material~\cite{supp}). 
Here we note: 
(i) the longitudinal entry numerically agrees with the density response though with a slightly different damping
model \cite{kallin_excitations_1984,giuliani_quantum_2005}; 
(ii) the Hall contribution arising from the time-reversal symmetry breaking imposed by the magnetic field mixes transverse and longitudinal electric fields, and relates to the Hall viscosity \cite{hoyos_hall_2012,harper_finite-wave-vector_2018};
(iii)
the {\it local} LL response \cite{palik_infrared_1970,wang_direct_2010,ismail_analytic_2016} is recovered from Eq.~\eqref{eq: LL nonlocal susceptibility tensor wavevector space} at \(q=0\), including the Drude susceptibility \cite{drude_zur_1900,landau_vibrations_1945} for $B=0$.

In the thin film approximation \cite{malic_graphene_2013,greten_microscopic_2025} (valid for a single QW), the electric field $\mathbf E(\mathbf q,z;\omega)$ emitted by a surface dipole density distribution $\mathbf P^{2\mathrm D}$ of an ultrathin layer is given by
\begin{align}
    \mathbf E(\mathbf q,z;\omega)
    =
    \frac{1}{\varepsilon_0}
    \mathbcal{G}^{(0)}(\mathbf q,z,z'=0;\omega)
    \cdot
    \mathbf P^{2\mathrm D}(\mathbf q;\omega),
    \label{eq:E_sheet_general}
\end{align}
where $\mathbcal{G}^{(0)}$ is the GF [units of inverse volume] of any layered environment excluding this ultrathin layer \cite{greten_dipolar_2024}.
Although interfaces could be readily introduced~\cite{salzwedel_spatial_2023,katzer_impact_2023}, we restrict to a homogeneous environment for clarity.
Self-consistently
solving Eq.~\eqref{eq:E_sheet_general} with Eq.~\eqref{eq: definition of chi} yields the scattered GF taking the ultrathin layer at $z=0$ into account.

For emission and observation at the same dipole position, $z_0\neq 0$, see Fig.~\ref{fig: intro pic}(b),
the scattered GF in the quasi-static limit ($c_0\rightarrow \infty$) \cite{greten_dipolar_2024,larsson_electromagnetics_2007}:
becomes
diagonal in the cartesian basis,
\begin{align}
\label{eq: Greens function, quasi-static}
&\mathbcal{G}^{\mathrm{sc}}_{\mathrm{qs}}(\boldsymbol{\rho}-\boldsymbol{\rho}'=\mathbf{0},z=z_0,z'=z_0;\omega)\\
&=
\int_0^\infty\!\!\!\!\!\!\mathrm{d}q\,
\frac{q^2\,
e^{-2q|z_0|}}{8\pi\varepsilon}
\frac{q\chi^L(q;\omega)}{ \left(2\varepsilon+q\chi^L(q;\omega)\right)}
\begin{pmatrix}
1 & 0 & 0 \\
0 & 1 & 0 \\
0 & 0 & 2
\end{pmatrix}. \nonumber
\end{align}
The complete
GF is shown (derived) in 
the End Matter (SM). 
Note that 
the near-field response does not distinguish between the co- and counter-rotating circular polarization directions.

We compute the Purcell factor---enhanced
emission factor from a dipole (such as an artificial atom quantum emitter)
at $\mathbf{r}_0 = (\mathbf{0},z_0)^\mathsf{T}$, with resonance frequency
$\omega_0$---as
$F_{\rm P} = 
\frac{6 \pi}{\sqrt{\varepsilon}\, {\omega_0^3}/{c_0}^3} 
\Im{
\mathbf{e}_{\sigma}\cdot
\mathbcal{G}^\text{sc}_{\mathrm{qs}}(\mathbf{0},z_0,z_0;\omega_0)
\cdot\mathbf{e}_{\sigma}}$
\cite{van_vlack_spontaneous_2012}
and the environment-induced Lamb shift,
$\Delta\omega
= 
-
\frac{d^2}{\hbar{\varepsilon_0}}
\Re{
\mathbf{e}_{\sigma}\cdot
\mathbcal{G}^\text{sc}_{\mathrm{qs}}(\mathbf{0},z_0,z_0;\omega_0)\cdot\mathbf{e}_{\sigma}},$
for an in-plane 
circular polarized dipole, i.e.,
$\mathbf{d}={d}\mathbf{e}_{\sigma}$ with $\mathbf{e}_{\sigma}=(1,i\sigma,0)^\mathsf{T}/\sqrt{2}$.

The dipole-emitted field spectrum, treated as a local oscillator
with resonance frequency $\omega_0$~\cite{deVries_point_1998}, detected vertically above (at $z_d$) is \cite{hughes_reconciling_2024}
\begin{equation}
S(z_0, z_d, \omega)
=
\left | \frac{\frac{2 \omega_0}{\hbar\varepsilon_0} 
\mathbf{d}\cdot
\mathbcal{G}(z_d,z_0;\omega)\cdot\mathbf{d}}{\omega_0^2-\omega^2 
- \frac{2 \omega_0}{\hbar\varepsilon_0} \mathbf{d}\cdot
\mathbcal{G}^{\text{sc}}_{\text{qs}}(z_0,z_0;\omega)\cdot\mathbf{d}}
\right |^2
,
\label{eq: spectrum}
\end{equation}
which 
agrees 
with a full quantum theory
of light emission from an excited
two level system~\cite{van_vlack_spontaneous_2012}
under linear response. Note that we have not made any rotating wave approximations, nor assumed weak light-matter coupling.
The same-point GF ${\mathbcal{G}}^{\rm sc}_{\rm qs}(z_0,z_0)$ in the denominator, incorporates Purcell factors and Lamb shifts,
and is a local quantity; however,
\textit{it includes LL nonlocality}.
The two-point GF ${\mathbcal{G}}(z_d,z_0)$ is
a nonlocal propagator
for far-field detection, $z_d>\lambda$, using a fully retarded form, but note that a local LL response is sufficient here; this latter function causes additional spectral reshaping, and accounts for light propagation effects from the dipole to the detector.

\begin{figure}[t] 
\centering
    \includegraphics[width=\linewidth]{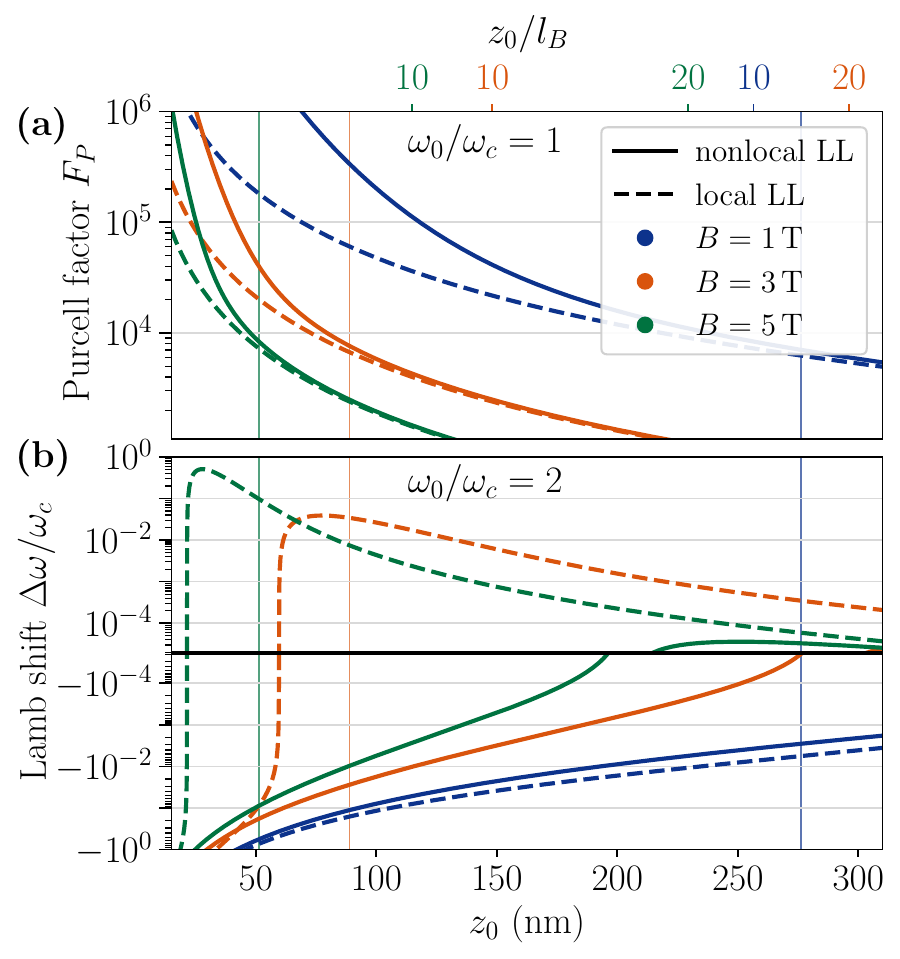}
    \caption{
    \textbf{(a)}\,/\,\textbf{(b)}: Purcell factor / Lamb shift for an in-plane dipole with $\omega_0 = \omega_c$ / $\omega_0 = 2 \omega_c$ in close proximity to a Landau-quantized 2DEG for magnetic field strengths $B=[1\,$(blue)$\, ,3\,$(orange)$\,,5\,$(green)$]\,$T.
    The vertical lines indicate twice the Larmor radius, $2r_c(\ell_F)$, of the highest occupied LL. Parameters: temperature $T=10\,$K, dipole moment $d = 100\,e$nm.
    } 
    \label{fig: qs Greens very near field}
\end{figure}

{\em Numerical Results.}---Figure~\ref{fig: qs Greens imag}(a) displays the imaginary 
and real (inset) part of the $q$-resolved susceptibility
[cf.~Eq.~\eqref{eq: LL nonlocal susceptibility tensor wavevector space}], using \(B=3\,\mathrm{T}\). At small \(q\), the response is dominated by the {\it dipole-allowed transition} at \(\omega=\omega_c\) (solid), while higher transitions, i.e., 
\(\omega=2\omega_c\) (dotted), are suppressed.
High-$q$ fields on a scale set by the inverse Larmor radius of the highest occupied LL, $r_c(\ell_F)$, can resolve the cyclotron motion, which reduces the \(\omega=\omega_c\) response {\it and activates dipole-forbidden transitions}, which are accessible to an emitter in the near-field.

The longitudinal (orange), transverse (red), and Hall (purple) components differ substantially, showing that the nonlocal LL response is not captured by a scalar correction to the local LL response.
At about the inverse magnetic length, \(q=l_B^{-1} = r_c^{-1}(\ell=0)\), the imaginary part of the longitudinal susceptibility drops to zero, suppressing electron density fluctuations below the smallest possible Larmor radius.
All parameters are given in the End Matter.

Fig.~\ref{fig: qs Greens imag}(b) shows the Purcell factor $F_P$
(as a measure of the projected local density of photon states) of an in-plane dipole above a Landau-quantized 2DEG for magnetic fields $B=1\,\mathrm{T}$ and $\,5\,\mathrm{T}$ and frequencies \(\omega_0/\omega_c=1,2,3\) at emitter-2DEG separations in the $\mu$m range corresponding to a few percent of the THz wavelength in the surrounding medium. 
An evanescent field carrying large in-plane wavevectors (strong field gradients),
enables Förster-type energy transfer beyond the dipole approximation to the 2DEG via
dipole-forbidden transitions at \(\omega_0=2\omega_c\) (orange) and \(\omega_0=3\omega_c\) (green). Consequently, 
{\it the nonlocal Purcell factor (solid) can exceed the local prediction (dashed) by up to two orders of magnitude for \(\omega_0=2\omega_c\)}, highlighted by vertical arrows.
This impact hundreds of nanometers away (well beyond the magnetic length $l_B$) arises because small variations of the electric field on the length scale of twice the highest occupied Larmor radius $2r_c(\ell_F)$, see Fig.~\ref{fig: qs Greens imag}a, suffice for significant deviations from the local LL approximation.

As evident from the strong deviations of the dashed lines between both panels of Fig.~\ref{fig: qs Greens imag}, the local response heavily depends on the absolute values of $z_0$ and $\omega_0$, so the local-nonlocal discrepancy grows with $B$.
However, the Purcell factor at the dipole-allowed transition, \(\omega_0=\omega_c\) (blue), does not deviate due to LL nonlocality.
Here, nonlocal LL corrections become relevant at distances comparable to the Larmor radius of occupied LL, see Fig.~\ref{fig: qs Greens very near field}(a). For $B\in[1,5]\,$T this is about $25$-$140\,$nm---a common length scale in modern photonic structures \cite{timmer_ultrafast_2026,andberger_terahertz_2024,kim_terahertz_2023}.
Due to longer magnetic lengths, \(l_B\sim B^{-1/2}\), deviations occur at larger distances for smaller \(B\). Also, $\ell_F$ decreases with increasing \(B\), such that only lower Landau levels (with smaller Larmor radii) are occupied.

On a similar length scale, but for $\omega_0=2\omega_c$ the Lamb shift can acquire the opposite sign compared to the local LL prediction, see Fig.~\ref{fig: qs Greens very near field}(b). Its magnitude reaches up to a considerable fraction of the cyclotron frequency $\omega_c$; again, local and nonlocal LL models describe qualitatively different scenarios: the nonlocal response opens a resonant coupling channel at $2\omega_c$, whereas the local response is determined by the off-resonant tail of the $\omega_c$ transition.

\begin{figure*}[t] 
\centering
    \includegraphics[width=1\linewidth]{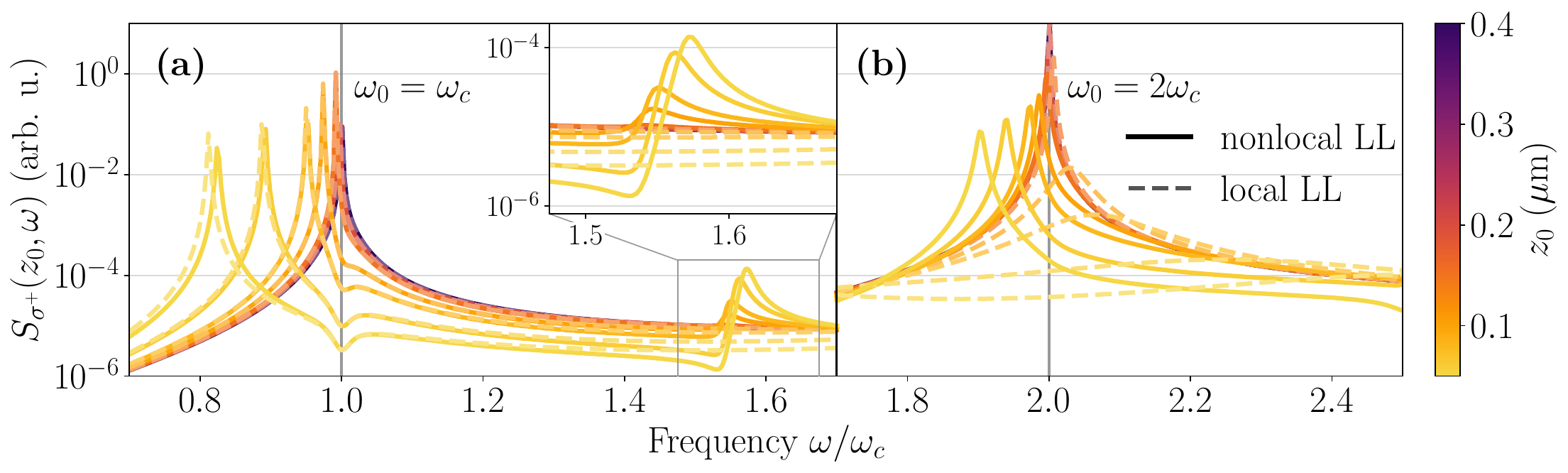}
    \caption{
    Emitted spectrum $S$ of an in-plane circular polarized point dipole with dipole moment $d=100\,e$nm and frequency \textbf{(a)} $\omega_0=\omega_c$ , and \textbf{(b)} $\omega_0=2\omega_c$,  indicated by grey vertical lines at $z_0$ above a Landau-quantized 2DEG (GaAs QW at temperature $T=10\,$K) for $B=3\,$T detected at $z_d = 5\,$mm (far-field) based on the nonlocal (solid) or local (dashed) LL response.
    The inset magnifies the collective magnetoplasmon resonance near $1.5$-$1.6\omega_c$, appearing only in the nonlocal LL response (solid) for small $z_0$.
    The discontinuity at the panel-intersection occurs since the panels depict spectra of dipoles with different resonance frequencies $\omega_0$.} 
    \label{fig: spectrum}
\end{figure*}

Figure~\ref{fig: spectrum} shows the spectrum $S(z_0,\omega)$ given by Eq.~\eqref{eq: spectrum}, emitted by a point dipole at $z_0$ in the near-field of the Landau-quantized 2DEG for $B=3\,$T and detected at $z_d = 5\,\mathrm{mm}$ in the far-field, i.e., multiples of the wavelength $\lambda$ away.
The gray vertical lines mark the bare dipole resonance frequency, $\omega_0 = \omega_c$ in (a) and $\omega_0 = 2\omega_c$ in (b), corresponding to wavelengths $\lambda(\varepsilon,\omega_0)$ of tens of $\mu$m.
In Fig.~\ref{fig: spectrum}(a) ($\omega_0 = \omega_c$), the dominant spectral feature appears close to $\omega_c$ for large $z_0$ (purple).
As $z_0$ decreases (toward yellow), the coupling to evanescent in-plane modes $\mathbf{q}$ of the 2DEG (i.e., Förster transfer) increases, leading to a pronounced reshaping and shift of the resonance.
Since the $\omega_c$ transition is present in both local and nonlocal LL model, the spectra remain qualitatively similar around the main cyclotron resonance.
A qualitative difference occurs at $\omega = 1.5-1.6 \omega_c$, arising from evanescent coupling to finite-wavevector LL excitations, so-called magneto-plasmons \cite{roldan_collective_2009,kallin_excitations_1984}:
The GF, Eq.~\eqref{eq: Greens function, quasi-static}, does not depend on the pure LL susceptibility as shown in Fig.~\ref{fig: qs Greens imag}(a), instead the screened susceptibility incorporates electron-electron interactions via the $(2\varepsilon+q\chi^L(q,\omega))$ denominator, here, featuring a (collective) resonance at $\omega = 1.5-1.6 \omega_c$, however, absent in the local LL approximation. This is a possible explanation for features observed for a subwavelength cavity-coupled 2DEG
in terahertz transmission
\cite{Rajabali_polaritonic_2021}.

Even stronger deviations from the local LL response occur for $\omega_0 = 2\omega_c$, see Fig.~\ref{fig: spectrum}(b).
Whereas the local response here constitutes an off-resonant interaction with the $\omega = \omega_c$ transition, the nonlocal spectrum resembles a resonant interaction. This leads to significant differences in linewidth (related to the Purcell factor) and a redshift (nonlocal LL) instead of a blueshift (local LL), cf.~Fig.~\ref{fig: qs Greens very near field}(b).

{\it Conclusions.}---We introduced a microscopic theory for the nonlocal susceptibility of a Landau-quantized 2DEG and used it to construct a self-consistent, fully retarded GF.
As an example, we demonstrated how the spatially inhomogeneous electric near-field of a single emitter relaxes the dipole selection rules and brightens locally forbidden LL transitions at multiples of the cyclotron frequency. We found that this brightening enhances the Purcell factor at $2\omega_c$ by up to two orders of magnitude relative to a local LL description out to distances of several hundreds of nanometers, and reverses the sign of the environment-induced Lamb shift at smaller separations; even the dipole-allowed cyclotron channel at $\omega_c$ acquires significant nonlocal corrections for distances up to $\approx100\,$nm (depending on QW material, environment and magnetic field strength).
Beyond these single-particle features directly arising from the finite spatial extent of the quantized cyclotron motion, our self-consistent treatment covers collective magnetoplasmon resonance, here appearing near $1.6\,\omega_c$, consistent with the spectral features observed in ultrastrongly coupled cavities (but so far unexplained)~\cite{Rajabali_polaritonic_2021}.

These results show that a nonlocal material model becomes indispensable when the electric field varies over the spatial extent of the occupied cyclotron orbits---a length scale reaching far beyond typical plasmonic nonlocality length scales below a few nanometers in the visible regime and well beyond the magnetic length $l_B$. Moreover, this scale coincides with those of state-of-the-art nanostructured THz-photonic architectures~\cite{kim_terahertz_2023,andberger_terahertz_2024,timmer_ultrafast_2026}.

\vspace{0.5cm}

{\it Acknowledgments.}---L.G. (and S.H) acknowledges support from the Alexander von Humboldt Foundation 
through a Feodor Lynen Research Fellowship
(Humboldt Award).
S.M. and A.K. acknowledge financial support from the Deutsche Forschungsgemeinschaft (DFG) through Project No.~556436549.
L.G and S.H. acknowledge financial support from the Natural Sciences and Engineering Research Council of Canada (NSERC). 
Finally, we thank Dasom Kim and Junichiro Kono for enlightening discussions and David Greten for numerical support in solving the Green's function Fourier integrals. We acknowledge the assistance of generative AI (ChatGPT and Claude) for plot generation and improving the language clarity of this manuscript.



\vspace{0.2cm}

{\bf Data availability.} The data that support the findings of this article are not publicly available upon publication. 
The data are available from the authors upon reasonable request.

\bibliography{bib}

@PREAMBLE{
 "\providecommand{\noopsort}[1]{}" 
 # "\providecommand{\singleletter}[1]{#1}%" 
}

@article{wu_strong_2019,
  title={Strong Purcell effect for terahertz magnetic dipole emission with spoof plasmonic structure},
  author={Wu, Hong-Wei and Li, Yang and Chen, Hua-Jun and Sheng, Zong-Qiang and Jing, Hao and Fan, Ren-Hao and Peng, Ru-Wen},
  journal={ACS Applied Nano Materials},
  volume={2},
  number={2},
  pages={1045--1052},
  year={2019},
  publisher={ACS Publications},
  doi = {10.1021/acsanm.8b02318}
}

@misc{supp,
  note = "See Supplemental Material at
    URL-will-be-inserted-by-publisher for the derivation of the nonlocal LL susceptibility and the Green's function."
}

@misc{RPA,
  note = "The Coulomb electron-electron interaction is commonly incorporated in the longitudinal susceptibility in the random phase approximation~\cite{kallin_excitations_1984,andolina_quantum_2026,giuliani_quantum_2005}. In our case, the same expression appears automatically via the self-consistent solution of Maxwell's equations." 
}

@article{joulain_surface_2005,
  title={Surface electromagnetic waves thermally excited: Radiative heat transfer, coherence properties and Casimir forces revisited in the near field},
  author={Joulain, Karl and Mulet, Jean-Philippe and Marquier, Fran{\c{c}}ois and Carminati, R{\'e}mi and Greffet, Jean-Jacques},
  journal={Surface Science Reports},
  volume={57},
  number={3-4},
  pages={59--112},
  year={2005},
  publisher={Elsevier},
  doi = {10.1016/j.surfrep.2004.12.002}
}

@article{li_purcell_2026,
	title = {Purcell enhancement of directional edge photocurrent in a van der {Waals} self-cavity},
	volume = {17},
	issn = {2041-1723},
	url = {https://www.nature.com/articles/s41467-026-72260-8},
	doi = {10.1038/s41467-026-72260-8},
	language = {en},
	number = {1},
	urldate = {2026-06-04},
	journal = {Nature Communications},
	author = {Li, Xinyu and Hagelstein, Jesse and Kipp, Gunda and Sturm, Felix and Kusyak, Kateryna and Huang, Yunfei and Schulte, Benedikt and Potts, Alexander M. and Stensberg, Jonathan and Quirós-Cordero, Victoria and Trovatello, Chiara and Peng, Zhi Hao and Hu, Chaowei and DeStefano, Jonathan M. and Fechner, Michael and Taniguchi, Takashi and Watanabe, Kenji and Schuck, P. James and Xu, Xiaodong and Chu, Jiun-Haw and Zhu, Xiaoyang and Rubio, Angel and Michael, Marios H. and Day, Matthew W. and Bretscher, Hope M. and McIver, James W.},
	month = apr,
	year = {2026},
	pages = {3865},
}

@article{todorov_purcell_2007,
  title = {Purcell Enhancement of Spontaneous Emission from Quantum Cascades inside Mirror-Grating Metal Cavities at THz Frequencies},
  author = {Todorov, Yanko and Sagnes, Isabelle and Abram, Izo and Minot, Christophe},
  journal = {Phys. Rev. Lett.},
  volume = {99},
  issue = {22},
  pages = {223603},
  numpages = {4},
  year = {2007},
  month = {Nov},
  publisher = {American Physical Society},
  doi = {10.1103/PhysRevLett.99.223603},
  url = {https://link.aps.org/doi/10.1103/PhysRevLett.99.223603}
}

@article{rathje_coupling_2023,
  title={Coupling broadband terahertz dipoles to microscale resonators},
  author={Rathje, Christopher and von Seggern, Rieke and Gräper, Leon A and Kredl, Jana and Walowski, Jakob and Münzenberg, Markus and Schäfer, Sascha},
  journal={ACS Photonics},
  volume={10},
  number={10},
  pages={3467--3475},
  year={2023},
  publisher={ACS Publications},
  doi = {10.1021/acsphotonics.3c00833}
}

@article{eriksen_nonlocal_2024,
  title={Nonlocal effects in plasmon-emitter interactions},
  author={Eriksen, Mikkel Have and Tserkezis, Christos and Mortensen, N Asger and Cox, Joel D},
  journal={Nanophotonics},
  volume={13},
  number={15},
  pages={2741--2751},
  year={2024},
  publisher={De Gruyter},
  doi = {10.1515/nanoph-2023-0575}
}

@article{Rajabali_polaritonic_2021,
  doi = {10.1038/s41566-021-00854-3},
  url = {https://doi.org/10.1038/s41566-021-00854-3},
  year = {2021},
  month = aug,
  publisher = {Springer Science and Business Media {LLC}},
  volume = {15},
  number = {9},
  pages = {690--695},
  author = {Shima Rajabali and Erika Cortese and Mattias Beck and Simone De Liberato and J{\'{e}}r{\^{o}}me Faist and Giacomo Scalari},
  title = {Polaritonic nonlocality in light{\textendash}matter interaction},
  journal = {Nature Photonics}
}

@article{novoselov_two_2005,
  title={Two-dimensional gas of massless Dirac fermions in graphene},
  author={Novoselov, Kostya S and Geim, Andre K and Morozov, Sergei Vladimirovich and Jiang, Dingde and Katsnelson, Michail I and Grigorieva, Irina V and Dubonos, Sergey V and Firsov, Alexandr A},
  journal={Nature},
  volume={438},
  number={7065},
  pages={197--200},
  year={2005},
  publisher={Nature Publishing Group UK London},
  doi = {10.1038/nature04233}
}

@article{laughlin_quantized_1981,
  title = {Quantized Hall conductivity in two dimensions},
  author = {Laughlin, R. B.},
  journal = {Phys. Rev. B},
  volume = {23},
  issue = {10},
  pages = {5632(R)--5633(R)},
  numpages = {0},
  year = {1981},
  month = {May},
  publisher = {American Physical Society},
  doi = {10.1103/PhysRevB.23.5632},
  url = {https://link.aps.org/doi/10.1103/PhysRevB.23.5632}
}

@article{li_vacuum_2018,
  title={Vacuum Bloch--Siegert shift in Landau polaritons with ultra-high cooperativity},
  author={Li, Xinwei and Bamba, Motoaki and Zhang, Qi and Fallahi, Saeed and Gardner, Geoff C and Gao, Weilu and Lou, Minhan and Yoshioka, Katsumasa and Manfra, Michael J and Kono, Junichiro},
  journal={Nature Photonics},
  volume={12},
  number={6},
  pages={324--329},
  year={2018},
  publisher={Nature Publishing Group UK London},
  doi = {10.1038/s41566-018-0153-0}
}

@article{zhang_superradiant_2014,
  title={Superradiant decay of cyclotron resonance of two-dimensional electron gases},
  author={Zhang, Qi and Arikawa, Takashi and Kato, Eiji and Reno, John L and Pan, Wei and Watson, John D and Manfra, Michael J and Zudov, Michael A and Tokman, Mikhail and Erukhimova, Maria and others},
  journal={Phys. Rev. Lett.},
  volume={113},
  number={4},
  pages={047601},
  year={2014},
  publisher={APS},
  doi = {10.1103/PhysRevLett.113.047601}
}

@article{jasper_broadband_2020,
  title={Broadband circular polarization time-domain terahertz spectroscopy},
  author={Jasper, Evan V and Mai, TT and Warren, MT and Smith, RK and Heligman, DM and McCormick, E and Ou, YS and Sheffield, M and Vald{\'e}s Aguilar, R},
  journal={Physical Review Materials},
  volume={4},
  number={1},
  pages={013803},
  year={2020},
  publisher={APS},
  doi = {10.1103/PhysRevMaterials.4.013803}
}

@article{ciraci_probing_2012,
  title={Probing the ultimate limits of plasmonic enhancement},
  author={Cirac{\`\i}, Cristian and Hill, RT and Mock, JJ and Urzhumov, Yaroslav and Fern{\'a}ndez-Dom{\'\i}nguez, AI and Maier, SA and Pendry, JB and Chilkoti, Ashutosh and Smith, DR},
  journal={Science},
  volume={337},
  number={6098},
  pages={1072--1074},
  year={2012},
  publisher={American Association for the Advancement of Science},
  doi = {10.1126/science.1224823}
}

@article{hagemuller_ultrastrong_2010,
  title = {Ultrastrong coupling between a cavity resonator and the cyclotron transition of a two-dimensional electron gas in the case of an integer filling factor},
  author = {Hagenm\"uller, David and De Liberato, Simone and Ciuti, Cristiano},
  journal = {Phys. Rev. B},
  volume = {81},
  issue = {23},
  pages = {235303},
  numpages = {9},
  year = {2010},
  month = {Jun},
  publisher = {American Physical Society},
  doi = {10.1103/PhysRevB.81.235303},
  url = {https://link.aps.org/doi/10.1103/PhysRevB.81.235303}
}

@article{palanen_quantized_1982,
  title = {Quantized Hall effect at low temperatures},
  author = {Paalanen, M. A. and Tsui, D. C. and Gossard, A. C.},
  journal = {Phys. Rev. B},
  volume = {25},
  issue = {8},
  pages = {5566(R)--5569(R)},
  numpages = {0},
  year = {1982},
  month = {Apr},
  publisher = {American Physical Society},
  doi = {10.1103/PhysRevB.25.5566},
  url = {https://link.aps.org/doi/10.1103/PhysRevB.25.5566}
}

@article{tsui_two_1982,
  title = {Two-Dimensional Magnetotransport in the Extreme Quantum Limit},
  author = {Tsui, D. C. and Stormer, H. L. and Gossard, A. C.},
  journal = {Phys. Rev. Lett.},
  volume = {48},
  issue = {22},
  pages = {1559--1562},
  numpages = {0},
  year = {1982},
  month = {May},
  publisher = {American Physical Society},
  doi = {10.1103/PhysRevLett.48.1559},
  url = {https://link.aps.org/doi/10.1103/PhysRevLett.48.1559}
}

@article{roldan_collective_2009,
  title = {Collective modes of doped graphene and a standard two-dimensional electron gas in a strong magnetic field: Linear magnetoplasmons versus magnetoexcitons},
  author = {Rold\'an, R. and Fuchs, J.-N. and Goerbig, M. O.},
  journal = {Phys. Rev. B},
  volume = {80},
  issue = {8},
  pages = {085408},
  numpages = {6},
  year = {2009},
  month = {Aug},
  publisher = {American Physical Society},
  doi = {10.1103/PhysRevB.80.085408},
  url = {https://link.aps.org/doi/10.1103/PhysRevB.80.085408}
}

@article{zhang_experimental_2005,
  title={Experimental observation of the quantum Hall effect and Berry's phase in graphene},
  author={Zhang, Yuanbo and Tan, Yan-Wen and Stormer, Horst L and Kim, Philip},
  journal={Nature},
  volume={438},
  number={7065},
  pages={201--204},
  year={2005},
  publisher={Nature Publishing Group UK London},
  doi = {10.1038/nature04235}
}

@article{landau_diamagnetismus_1930,
  title={Diamagnetismus der {M}etalle [{D}iamagnetism of metals]},
  author={Landau, LD},
  journal={Zeitschrift f{\"u}r Physik},
  volume={64},
  number={9},
  pages={629--637},
  year={1930},
  publisher={Springer},
  doi = {10.1007/BF01397213}
}

@article{rajabali_ultrastrongly_2022,
	title = {An ultrastrongly coupled single terahertz meta-atom},
	volume = {13},
	issn = {2041-1723},
	url = {https://www.nature.com/articles/s41467-022-29974-2},
	doi = {10.1038/s41467-022-29974-2},
	language = {en},
	number = {1},
	urldate = {2026-06-26},
	journal = {Nature Communications},
	author = {Rajabali, Shima and Markmann, Sergej and Jöchl, Elsa and Beck, Mattias and Lehner, Christian A. and Wegscheider, Werner and Faist, Jérôme and Scalari, Giacomo},
	month = may,
	year = {2022},
	pages = {2528},
}

@article{andolina_quantum_2026,
	title = {Quantum electrodynamics of graphene {Landau} levels in a deep-subwavelength hyperbolic phonon-polariton cavity},
	volume = {8},
	issn = {2643-1564},
	url = {https://link.aps.org/doi/10.1103/69jg-sq8r},
	doi = {10.1103/69jg-sq8r},
	language = {en},
	number = {2},
	urldate = {2026-06-26},
	journal = {Physical Review Research},
	author = {Andolina, Gian Marcello and Ceccanti, Matteo and Turini, Bianca and Riolo, Riccardo and Polini, Marco and Schirò, Marco and Koppens, Frank H. L.},
	month = apr,
	year = {2026},
	pages = {023044},
}

@article{endo_cavitymediated_2025,
	title = {Cavity‐mediated coupling between local and nonlocal modes in {Landau} polaritons},
	volume = {14},
	copyright = {http://creativecommons.org/licenses/by/4.0/},
	issn = {2192-8614, 2192-8614},
	url = {https://onlinelibrary.wiley.com/doi/10.1515/nanoph-2025-0442},
	doi = {10.1515/nanoph-2025-0442},
	language = {en},
	number = {25},
	urldate = {2026-06-26},
	journal = {Nanophotonics},
	author = {Endo, Sae R. and Kim, Dasom and Liang, Shuang and Lee, Geon and Kim, Sunghwan and Covarrubias‐Morales, Alan and Seo, Minah and Manfra, Michael J. and Lee, Dukhyung and Bamba, Motoaki and Kono, Junichiro},
	month = dec,
	year = {2025},
	pages = {4647--4654},
}

@article{yahniuk_strongly_2026,
	title = {Strongly nonlinear {Bernstein} modes in graphene reveal plasmon-enhanced near-field magnetoabsorption},
	volume = {113},
	issn = {2469-9950, 2469-9969},
	url = {https://link.aps.org/doi/10.1103/32bf-v28g},
	doi = {10.1103/32bf-v28g},
	language = {en},
	number = {12},
	urldate = {2026-06-26},
	journal = {Physical Review B},
	author = {Yahniuk, I. and Dmitriev, I. A. and Shilov, A. L. and Mönch, E. and Marocko, M. and Eroms, J. and Weiss, D. and Sadovyi, P. and Sadovyi, B. and Grzegory, I. and Knap, W. and Gumenjuk-Sichevska, J. and Wunderlich, J. and Bandurin, D. A. and Ganichev, S. D.},
	month = mar,
	year = {2026},
	pages = {125418},
}

@article{scalari_ultrastrong_2012,
	title = {Ultrastrong {Coupling} of the {Cyclotron} {Transition} of a {2D} {Electron} {Gas} to a {THz} {Metamaterial}},
	volume = {335},
	issn = {0036-8075, 1095-9203},
	url = {https://www.science.org/doi/10.1126/science.1216022},
	doi = {10.1126/science.1216022},
	language = {en},
	number = {6074},
	urldate = {2026-06-26},
	journal = {Science},
	author = {Scalari, G. and Maissen, C. and Turčinková, D. and Hagenmüller, D. and De Liberato, S. and Ciuti, C. and Reichl, C. and Schuh, D. and Wegscheider, W. and Beck, M. and Faist, J.},
	month = mar,
	year = {2012},
	pages = {1323--1326},
}

@article{deVries_point_1998,
  title = {Point scatterers for classical waves},
  author = {de Vries, Pedro and van Coevorden, David V. and Lagendijk, Ad},
  journal = {Rev. Mod. Phys.},
  volume = {70},
  issue = {2},
  pages = {447--466},
  numpages = {0},
  year = {1998},
  month = {Apr},
  publisher = {American Physical Society},
  doi = {10.1103/RevModPhys.70.447},
  url = {https://link.aps.org/doi/10.1103/RevModPhys.70.447}
}

@article{larsson_electromagnetics_2007,
  title={Electromagnetics from a quasistatic perspective},
  author={Larsson, Jonas},
  journal={American Journal of Physics},
  volume={75},
  number={3},
  pages={230--239},
  year={2007},
}

@article{andberger_terahertz_2024,
	title = {Terahertz chiral subwavelength cavities breaking time-reversal symmetry via ultrastrong light-matter interaction},
	volume = {109},
	issn = {2469-9950, 2469-9969},
	url = {https://link.aps.org/doi/10.1103/PhysRevB.109.L161302},
	doi = {10.1103/PhysRevB.109.L161302},
	language = {en},
	number = {16},
	urldate = {2024-12-09},
	journal = {Physical Review B},
	author = {Andberger, Johan and Graziotto, Lorenzo and Sacchi, Luca and Beck, Mattias and Scalari, Giacomo and Faist, Jérôme},
	month = apr,
	year = {2024},
	pages = {L161302},
}

@article{timmer_ultrafast_2026,
	title = {Ultrafast transition from coherent to incoherent polariton nonlinearities in a hybrid {1L}-{WS2}/plasmon structure},
	volume = {21},
	issn = {1748-3387, 1748-3395},
	url = {https://www.nature.com/articles/s41565-025-02054-4},
	doi = {10.1038/s41565-025-02054-4},
	language = {en},
	number = {2},
	urldate = {2026-06-23},
	journal = {Nature Nanotechnology},
	author = {Timmer, Daniel and Gittinger, Moritz and Quenzel, Thomas and Cadore, Alisson R. and Rosa, Barbara L. T. and Li, Wenshan and Soavi, Giancarlo and Lünemann, Daniel C. and Stephan, Sven and Greten, Lara and Richter, Marten and Knorr, Andreas and De Sio, Antonietta and Silies, Martin and Cerullo, Giulio and Ferrari, Andrea C. and Lienau, Christoph},
	month = feb,
	year = {2026},
	pages = {216--222},
}

@article{hughes_reconciling_2024,
	title = {Reconciling quantum and classical spectral theories of ultrastrong coupling: role of cavity bath coupling and gauge corrections},
	volume = {2},
	copyright = {© 2024 Optica Publishing Group},
	issn = {2837-6714},
	shorttitle = {Reconciling quantum and classical spectral theories of ultrastrong coupling},
	url = {https://opg.optica.org/opticaq/abstract.cfm?uri=opticaq-2-3-133},
	doi = {10.1364/OPTICAQ.519395},
	language = {EN},
	number = {3},
	urldate = {2024-10-21},
	journal = {Optica Quantum},
	author = {Hughes, Stephen and Gustin, Chris and Nori, Franco},
	month = jun,
	year = {2024},
	pages = {133--139},
}

@article{richards_inelastic_2000,
	title = {Inelastic light scattering from inter-{Landau} level excitations in a two-dimensional electron gas},
	volume = {61},
	url = {https://link.aps.org/doi/10.1103/PhysRevB.61.7517},
	doi = {10.1103/PhysRevB.61.7517},
	number = {11},
	urldate = {2025-02-11},
	journal = {Physical Review B},
	author = {Richards, David},
	month = mar,
	year = {2000},
	pages = {7517--7525},
}

@article{oji_magnetoplasma_1986,
	title = {Magnetoplasma modes of the two-dimensional electron gas at nonintegral filling factors},
	volume = {33},
	copyright = {http://link.aps.org/licenses/aps-default-license},
	issn = {0163-1829},
	url = {https://link.aps.org/doi/10.1103/PhysRevB.33.3810},
	doi = {10.1103/PhysRevB.33.3810},
	language = {en},
	number = {6},
	urldate = {2025-02-18},
	journal = {Physical Review B},
	author = {Oji, H. C. A. and MacDonald, A. H.},
	month = mar,
	year = {1986},
	pages = {3810--3818},
}

@article{greten_dipolar_2024,
	title = {Dipolar {Coupling} at {Interfaces} of {Ultrathin} {Semiconductors}, {Semimetals}, {Plasmonic} {Nanoparticles}, and {Molecules}},
	volume = {221},
	issn = {1862-6319},
	url = {https://onlinelibrary.wiley.com/doi/abs/10.1002/pssa.202300102},
	doi = {10.1002/pssa.202300102},
	language = {en},
	number = {1},
	journal = {physica status solidi (a)},
	author = {Greten, Lara and Salzwedel, Robert and Katzer, Manuel and Mittenzwey, Henry and Christiansen, Dominik and Knorr, Andreas and Selig, Malte},
	year = {2024},
	pages = {2300102},
}

@article{raza_nonlocal_2015,
	title = {Nonlocal optical response in metallic nanostructures},
	volume = {27},
	issn = {0953-8984},
	url = {https://dx.doi.org/10.1088/0953-8984/27/18/183204},
	doi = {10.1088/0953-8984/27/18/183204},
	language = {en},
	number = {18},
	urldate = {2025-07-17},
	journal = {Journal of Physics: Condensed Matter},
	author = {Raza, Søren and Bozhevolnyi, Sergey I and Wubs, Martijn and Asger Mortensen, N},
	month = apr,
	year = {2015},
	pages = {183204},
}

@article{landau_vibrations_1945,
  title={On the vibrations of the electronic plasma},
  author={Landau, L},
  journal={Zhurnal eksperimentalnoi i teoreticheskoi fiziki},
  volume={16},
  number={7},
  pages={574--586},
  year={1946},
}

@article{forn-diaz_ultrastrong_2019,
	title = {Ultrastrong coupling regimes of light-matter interaction},
	volume = {91},
	issn = {0034-6861, 1539-0756},
	url = {https://link.aps.org/doi/10.1103/RevModPhys.91.025005},
	doi = {10.1103/RevModPhys.91.025005},
	number = {2},
	urldate = {2023-04-19},
	journal = {Reviews of Modern Physics},
	author = {Forn-Díaz, P. and Lamata, L. and Rico, E. and Kono, J. and Solano, E.},
	month = jun,
	year = {2019},
	pages = {025005},
}

@article{frisk_kockum_ultrastrong_2019,
	title = {Ultrastrong coupling between light and matter},
	volume = {1},
	issn = {2522-5820},
	url = {https://www.nature.com/articles/s42254-018-0006-2},
	doi = {10.1038/s42254-018-0006-2},
	number = {1},
	urldate = {2023-04-19},
	journal = {Nature Reviews Physics},
	author = {Frisk Kockum, Anton and Miranowicz, Adam and De Liberato, Simone and Savasta, Salvatore and Nori, Franco},
	month = jan,
	year = {2019},
	pages = {19--40},
}

@article{Yao_ultrahigh_2009,
  title = {Ultrahigh Purcell factors and Lamb shifts in slow-light metamaterial waveguides},
  author = {Yao, Peijun and Van Vlack, C. and Reza, A. and Patterson, M. and Dignam, M. M. and Hughes, S.},
  journal = {Phys. Rev. B},
  volume = {80},
  issue = {19},
  pages = {195106},
  numpages = {11},
  year = {2009},
  month = {Nov},
  publisher = {American Physical Society},
  doi = {10.1103/PhysRevB.80.195106},
  url = {https://link.aps.org/doi/10.1103/PhysRevB.80.195106}
}

@article{van_vlack_spontaneous_2012,
	title = {Spontaneous emission spectra and quantum light-matter interactions from a strongly coupled quantum dot metal-nanoparticle system},
	volume = {85},
	issn = {1098-0121, 1550-235X},
	url = {https://link.aps.org/doi/10.1103/PhysRevB.85.075303},
	doi = {10.1103/PhysRevB.85.075303},
	number = {7},
	urldate = {2023-04-12},
	journal = {Physical Review B},
	author = {Van Vlack, C. and Kristensen, Philip Trøst and Hughes, S.},
	month = feb,
	year = {2012},
	pages = {075303},
}

@article{salzwedel_spatial_2023,
  title = {Spatial exciton localization at interfaces of metal nanoparticles and atomically thin semiconductors},
  author = {Salzwedel, Robert and Greten, Lara and Schmidt, Stefan and Hughes, Stephen and Knorr, Andreas and Selig, Malte},
  journal = {Phys. Rev. B},
  volume = {109},
  issue = {3},
  pages = {035309},
  numpages = {16},
  year = {2024},
  month = jan,
  publisher = {American Physical Society},
  doi = {10.1103/PhysRevB.109.035309},
  url = {https://link.aps.org/doi/10.1103/PhysRevB.109.035309}
}

@article{ismail_analytic_2016,
	title = {Analytic properties of complex {Hermite} polynomials},
	volume = {368},
	issn = {0002-9947, 1088-6850},
	url = {https://www.ams.org/tran/2016-368-02/S0002-9947-2015-06358-2/},
	doi = {10.1090/tran/6358},
	language = {en},
	number = {2},
	urldate = {2025-07-03},
	journal = {Transactions of the American Mathematical Society},
	author = {Ismail, Mourad},
	month = feb,
	year = {2016},
	pages = {1189--1210},
}

@book{vogel_quantum_2006,
	title = {Quantum {Optics}},
	isbn = {978-3-527-60845-4},
	language = {en},
	publisher = {John Wiley \& Sons},
	author = {Vogel, Werner and Welsch, Dirk-Gunnar},
	month = aug,
    url = {https://www.wiley-vch.de/de?option=com_eshop&view=product&isbn=9783527405077&title=Quantum%20Optics}, 
	year = {2006},
}

@article{jain_composite_2015,
	title = {Composite {Fermion} {Theory} of {Exotic} {Fractional} {Quantum} {Hall} {Effect}},
	volume = {6},
	issn = {1947-5454, 1947-5462},
	url = {https://www.annualreviews.org/doi/10.1146/annurev-conmatphys-031214-014606},
	doi = {10.1146/annurev-conmatphys-031214-014606},
	language = {en},
	number = {1},
	urldate = {2025-02-26},
	journal = {Annual Review of Condensed Matter Physics},
	author = {Jain, Jainendra K.},
	month = mar,
	year = {2015},
	pages = {39--62},
}

@book{haken_quantenfeldtheorie_1993,
	address = {Wiesbaden},
	title = {Quantenfeldtheorie des {Festkörpers} [{Q}uantum Field Theory of Solids: An Introduction]},
	copyright = {http://www.springer.com/tdm},
	isbn = {978-3-519-13025-3},
	url = {http://link.springer.com/10.1007/978-3-322-99250-5},
	urldate = {2025-01-06},
	publisher = {Vieweg+Teubner Verlag},
	author = {Haken, Hermann},
	year = {1993},
	doi = {10.1007/978-3-322-99250-5},
}

@article{vagner_ideally_1983,
	title = {Ideally {Conducting} {Phases} in {Quasi} {Two}-{Dimensional} {Conductors}},
	volume = {51},
	url = {https://link.aps.org/doi/10.1103/PhysRevLett.51.1700},
	doi = {10.1103/PhysRevLett.51.1700},
	number = {18},
	urldate = {2025-02-11},
	journal = {Phys. Rev. Lett.},
	author = {Vagner, I. D. and Maniv, Tsofar and Ehrenfreund, E.},
	month = oct,
	year = {1983},
	pages = {1700--1703},
}

@article{greten_microscopic_2025,
	title = {Microscopic theory for a minimal oscillator model of exciton-plasmon coupling in hybrids of two-dimensional semiconductors and metal nanoparticles},
	volume = {111},
	url = {https://link.aps.org/doi/10.1103/PhysRevB.111.205438},
	doi = {10.1103/PhysRevB.111.205438},
	number = {20},
	urldate = {2025-05-28},
	journal = {Physical Review B},
	author = {Greten, Lara and Salzwedel, Robert and Schutsch, Diana and Knorr, Andreas},
	month = may,
	year = {2025},
	pages = {205438},
}

@article{palik_infrared_1970,
	title = {Infrared and microwave magnetoplasma effects in semiconductors},
	volume = {33},
	issn = {0034-4885},
	url = {https://dx.doi.org/10.1088/0034-4885/33/3/307},
	doi = {10.1088/0034-4885/33/3/307},
	language = {en},
	number = {3},
	urldate = {2025-01-09},
	journal = {Reports on Progress in Physics},
	author = {Palik, E. D. and Furdyna, J. K.},
	month = sep,
	year = {1970},
	pages = {1193},
}

@article{mortensen_generalized_2014,
	title = {A generalized non-local optical response theory for plasmonic nanostructures},
	volume = {5},
	copyright = {2014 Springer Nature Limited},
	issn = {2041-1723},
	url = {https://www.nature.com/articles/ncomms4809},
	doi = {10.1038/ncomms4809},
	language = {en},
	number = {1},
	urldate = {2025-05-06},
	journal = {Nature Communications},
	author = {Mortensen, N. A. and Raza, S. and Wubs, M. and Søndergaard, T. and Bozhevolnyi, S. I.},
	month = may,
	year = {2014},
	pages = {3809},
}

@book{giuliani_quantum_2005,
  place={Cambridge},
  title={Quantum Theory of the Electron Liquid}, 
  publisher={Cambridge University Press},
  author={Giuliani, Gabriele and Vignale, Giovanni},
  year={2005},
  doi = {10.1017/CBO9780511619915}
}

@article{paravicini-bagliani_magneto-transport_2019,
	title = {Magneto-transport controlled by {Landau} polariton states},
	volume = {15},
	issn = {1745-2473, 1745-2481},
	url = {https://www.nature.com/articles/s41567-018-0346-y},
	doi = {10.1038/s41567-018-0346-y},
	language = {en},
	number = {2},
	urldate = {2026-06-26},
	journal = {Nature Physics},
	author = {Paravicini-Bagliani, Gian L. and Appugliese, Felice and Richter, Eli and Valmorra, Federico and Keller, Janine and Beck, Mattias and Bartolo, Nicola and Rössler, Clemens and Ihn, Thomas and Ensslin, Klaus and Ciuti, Cristiano and Scalari, Giacomo and Faist, Jérôme},
	month = feb,
	year = {2019},
	pages = {186--190},
}

@article{muniz_casimir_2021,
	title = {Casimir forces in the flatland: {Interplay} between photoinduced phase transitions and quantum {Hall} physics},
	volume = {3},
	issn = {2643-1564},
	shorttitle = {Casimir forces in the flatland},
	url = {https://link.aps.org/doi/10.1103/PhysRevResearch.3.023061},
	doi = {10.1103/PhysRevResearch.3.023061},
	language = {en},
	number = {2},
	urldate = {2026-06-26},
	journal = {Physical Review Research},
	author = {Muniz, Y. and Farina, C. and Kort-Kamp, W. J. M.},
	month = apr,
	year = {2021},
	pages = {023061},
}

@article{cysne_tuning_2014,
	title = {Tuning the {Casimir}-{Polder} interaction via magneto-optical effects in graphene},
	volume = {90},
	copyright = {http://link.aps.org/licenses/aps-default-license},
	issn = {1050-2947, 1094-1622},
	url = {https://link.aps.org/doi/10.1103/PhysRevA.90.052511},
	doi = {10.1103/PhysRevA.90.052511},
	language = {en},
	number = {5},
	urldate = {2026-06-26},
	journal = {Physical Review A},
	author = {Cysne, T. and Kort-Kamp, W. J. M. and Oliver, D. and Pinheiro, F. A. and Rosa, F. S. S. and Farina, C.},
	month = nov,
	year = {2014},
	pages = {052511},
}

@article{wu_active_2019,
	title = {Active {Magneto}-{Optical} {Control} of {Near}-{Field} {Radiative} {Heat} {Transfer} between {Graphene} {Sheets}},
	volume = {11},
	issn = {2331-7019},
	url = {https://link.aps.org/doi/10.1103/PhysRevApplied.11.054020},
	doi = {10.1103/PhysRevApplied.11.054020},
	language = {en},
	number = {5},
	urldate = {2026-06-26},
	journal = {Physical Review Applied},
	author = {Wu, Huihai and Huang, Yong and Cui, Longji and Zhu, Keyong},
	month = may,
	year = {2019},
	pages = {054020},
}

@article{ma_local_2022,
	title = {Local and nonlocal {Purcell} factor control of an emitter in graphene under the modulation of a static magnetic field},
	volume = {39},
	issn = {0740-3224, 1520-8540},
	url = {https://opg.optica.org/abstract.cfm?URI=josab-39-2-556},
	doi = {10.1364/JOSAB.442888},
	language = {en},
	number = {2},
	urldate = {2026-06-26},
	journal = {Journal of the Optical Society of America B},
	author = {Ma, Zenghong and Chen, Zijian and Zhang, Lian and Lu, Xiaocui and Xu, Jian and Xu, Xin and Yang, Guangwu},
	month = feb,
	year = {2022},
	pages = {556},
}

@article{eriksen_chiral_2025,
  title = {Chiral near-field control of quantum light generation using magneto-optical graphene},
  author = {Eriksen, Mikkel Have and Cox, Joel D.},
  journal = {Phys. Rev. B},
  volume = {112},
  issue = {12},
  pages = {125429},
  numpages = {15},
  year = {2025},
  month = {Sep},
  publisher = {American Physical Society},
  doi = {10.1103/k5sp-pmcy},
  url = {https://link.aps.org/doi/10.1103/k5sp-pmcy}
}

@article{bandurin_cyclotron_2022,
	title = {Cyclotron resonance overtones and near-field magnetoabsorption via terahertz {Bernstein} modes in graphene},
	volume = {18},
	issn = {1745-2473, 1745-2481},
	url = {https://www.nature.com/articles/s41567-021-01494-8},
	doi = {10.1038/s41567-021-01494-8},
	language = {en},
	number = {4},
	urldate = {2026-06-26},
	journal = {Nature Physics},
	author = {Bandurin, D. A. and Mönch, E. and Kapralov, K. and Phinney, I. Y. and Lindner, K. and Liu, S. and Edgar, J. H. and Dmitriev, I. A. and Jarillo-Herrero, P. and Svintsov, D. and Ganichev, S. D.},
	month = apr,
	year = {2022},
	pages = {462--467},
}

@article{bialek_photoresponse_2015,
	title = {Photoresponse of a two-dimensional electron gas at the second harmonic of the cyclotron resonance},
	volume = {91},
	copyright = {http://link.aps.org/licenses/aps-default-license},
	issn = {1098-0121, 1550-235X},
	url = {https://link.aps.org/doi/10.1103/PhysRevB.91.045437},
	doi = {10.1103/PhysRevB.91.045437},
	language = {en},
	number = {4},
	urldate = {2026-06-26},
	journal = {Physical Review B},
	author = {Białek, M. and Łusakowski, J. and Czapkiewicz, M. and Wróbel, J. and Umansky, V.},
	month = jan,
	year = {2015},
	pages = {045437},
}

@article{volkov_bernstein_2014,
	title = {Bernstein modes and giant microwave response of a two-dimensional electron system},
	volume = {89},
	copyright = {http://link.aps.org/licenses/aps-default-license},
	issn = {1098-0121, 1550-235X},
	url = {https://link.aps.org/doi/10.1103/PhysRevB.89.121410},
	doi = {10.1103/PhysRevB.89.121410},
	language = {en},
	number = {12},
	urldate = {2026-06-26},
	journal = {Physical Review B},
	author = {Volkov, V. A. and Zabolotnykh, A. A.},
	month = mar,
	year = {2014},
	pages = {121410}
}

@article{maag_coherent_2016,
	title = {Coherent cyclotron motion beyond {Kohn}’s theorem},
	volume = {12},
	issn = {1745-2473, 1745-2481},
	url = {https://www.nature.com/articles/nphys3559},
	doi = {10.1038/nphys3559},
	language = {en},
	number = {2},
	urldate = {2026-06-26},
	journal = {Nature Physics},
	author = {Maag, T. and Bayer, A. and Baierl, S. and Hohenleutner, M. and Korn, T. and Schüller, C. and Schuh, D. and Bougeard, D. and Lange, C. and Huber, R. and Mootz, M. and Sipe, J. E. and Koch, S. W. and Kira, M.},
	month = feb,
	year = {2016},
	pages = {119--123},
}

@article{harper_finite-wave-vector_2018,
	title = {Finite-wave-vector electromagnetic response in lattice quantum {Hall} systems},
	volume = {98},
	issn = {2469-9950, 2469-9969},
	url = {https://link.aps.org/doi/10.1103/PhysRevB.98.245303},
	doi = {10.1103/PhysRevB.98.245303},
	language = {en},
	number = {24},
	urldate = {2026-06-26},
	journal = {Physical Review B},
	author = {Harper, Fenner and Bauer, David and Jackson, T. S. and Roy, Rahul},
	month = dec,
	year = {2018},
	pages = {245303},
}

@article{hoyos_hall_2012,
	title = {Hall {Viscosity} and {Electromagnetic} {Response}},
	volume = {108},
	copyright = {http://link.aps.org/licenses/aps-default-license},
	issn = {0031-9007, 1079-7114},
	url = {https://link.aps.org/doi/10.1103/PhysRevLett.108.066805},
	doi = {10.1103/PhysRevLett.108.066805},
	language = {en},
	number = {6},
	urldate = {2026-06-26},
	journal = {Phys. Rev. Lett.},
	author = {Hoyos, Carlos and Son, Dam Thanh},
	month = feb,
	year = {2012},
	pages = {066805},
}

@article{kukushkin_dispersion_2009,
	title = {Dispersion of the {Excitations} of {Fractional} {Quantum} {Hall} {States}},
	volume = {324},
	issn = {0036-8075, 1095-9203},
	url = {https://www.science.org/doi/10.1126/science.1171472},
	doi = {10.1126/science.1171472},
	number = {5930},
	urldate = {2026-06-26},
	journal = {Science},
	author = {Kukushkin, Igor V. and Smet, Jurgen H. and Scarola, Vito W. and Umansky, Vladimir and Von Klitzing, Klaus},
	month = may,
	year = {2009},
	pages = {1044--1047},
}

@article{pinczuk_observation_1988,
  title = {Observation of roton density of states in two-dimensional Landau-level excitations},
  author = {Pinczuk, A. and Valladares, J. P. and Heiman, D. and Gossard, A. C. and English, J. H. and Tu, C. W. and Pfeiffer, L. and West, K.},
  journal = {Phys. Rev. Lett.},
  volume = {61},
  issue = {23},
  pages = {2701--2704},
  numpages = {0},
  year = {1988},
  month = {Dec},
  publisher = {American Physical Society},
  doi = {10.1103/PhysRevLett.61.2701},
  url = {https://link.aps.org/doi/10.1103/PhysRevLett.61.2701}
}

@article{kallin_many-body_1985,
	title = {Many-body effects on the cyclotron resonance in a two-dimensional electron gas},
	volume = {31},
	copyright = {http://link.aps.org/licenses/aps-default-license},
	issn = {0163-1829},
	url = {https://link.aps.org/doi/10.1103/PhysRevB.31.3635},
	doi = {10.1103/PhysRevB.31.3635},
	language = {en},
	number = {6},
	urldate = {2026-06-26},
	journal = {Physical Review B},
	author = {Kallin, C. and Halperin, B. I.},
	month = mar,
	year = {1985},
	pages = {3635--3647},
}

@article{meyer_proposal_2026,
	title = {Proposal for resolving quantized {Landau} orbits via elastic {XUV} scattering},
	volume = {113},
	issn = {2469-9950, 2469-9969},
	url = {https://link.aps.org/doi/10.1103/hq91-75kb},
	doi = {10.1103/hq91-75kb},
	language = {en},
	number = {16},
	urldate = {2026-06-29},
	journal = {Physical Review B},
	author = {Meyer, Sabrina and Sturm, Joris and Schröder, Christina and Hughes, Stephen and Knorr, Andreas and Greten, Lara},
	month = apr,
	year = {2026},
	pages = {165421},
}

@article{macdonald_magnetoplasmon_1985,
	title = {Magnetoplasmon {Excitations} from {Partially} {Filled} {Landau} {Levels} in {Two} {Dimensions}},
	volume = {55},
	copyright = {http://link.aps.org/licenses/aps-default-license},
	issn = {0031-9007},
	url = {https://link.aps.org/doi/10.1103/PhysRevLett.55.2208},
	doi = {10.1103/PhysRevLett.55.2208},
	language = {en},
	number = {20},
	urldate = {2026-06-26},
	journal = {Phys. Rev. Lett.},
	author = {MacDonald, A. H. and Oji, H. C. A. and Girvin, S. M.},
	month = nov,
	year = {1985},
	pages = {2208--2211},
}

@article{girvin_magneto_1986,
  title = {Magneto-roton theory of collective excitations in the fractional quantum Hall effect},
  author = {Girvin, S. M. and MacDonald, A. H. and Platzman, P. M.},
  journal = {Phys. Rev. B},
  volume = {33},
  issue = {4},
  pages = {2481--2494},
  numpages = {0},
  year = {1986},
  month = {Feb},
  publisher = {American Physical Society},
  doi = {10.1103/PhysRevB.33.2481},
  url = {https://link.aps.org/doi/10.1103/PhysRevB.33.2481}
}

@article{kallin_excitations_1984,
	title = {Excitations from a filled {Landau} level in the two-dimensional electron gas},
	volume = {30},
	copyright = {http://link.aps.org/licenses/aps-default-license},
	issn = {0163-1829},
	url = {https://link.aps.org/doi/10.1103/PhysRevB.30.5655},
	doi = {10.1103/PhysRevB.30.5655},
	language = {en},
	number = {10},
	urldate = {2026-06-26},
	journal = {Physical Review B},
	author = {Kallin, C. and Halperin, B. I.},
	month = nov,
	year = {1984},
	pages = {5655--5668},
}

@article{chiu_plasma_1974,
  title={Plasma oscillations of a two-dimensional electron gas in a strong magnetic field},
  author={Chiu, KW and Quinn, JJ},
  journal={Physical Review B},
  volume={9},
  number={11},
  pages={4724},
  year={1974},
  doi = {10.1103/PhysRevB.9.4724}
}

@article{greene_linear_1969,
	title = {Linear {Response} {Theory} for a {Degenerate} {Electron} {Gas} in a {Strong} {Magnetic} {Field}},
	volume = {177},
	copyright = {http://link.aps.org/licenses/aps-default-license},
	issn = {0031-899X},
	url = {https://link.aps.org/doi/10.1103/PhysRev.177.1019},
	doi = {10.1103/PhysRev.177.1019},
	language = {en},
	number = {3},
	urldate = {2026-06-29},
	journal = {Physical Review},
	author = {Greene, Michael P. and Lee, Hyung Joon and Quinn, J. J. and Rodriguez, Sergio},
	month = jan,
	year = {1969},
	pages = {1019--1036},
}

@article{messelot_large_2022,
  title = {Large terahertz electric dipole of a single graphene quantum dot},
  author = {Messelot, Simon and Riccardi, Elisa and Massabeau, Sylvain and Valmorra, Federico and Rosticher, Michael and Watanabe, Kenji and Taniguchi, Takashi and Tignon, J\'er\^ome and Boulier, Thomas and Dauvois, Vincent and Delbecq, Matthieu and Dhillon, Sukhdeep and Balibar, S\'ebastien and Kontos, Takis and Mangeney, Juliette},
  journal = {Phys. Rev. Res.},
  volume = {4},
  issue = {1},
  pages = {L012018},
  numpages = {6},
  year = {2022},
  month = {Feb},
  publisher = {American Physical Society},
  doi = {10.1103/PhysRevResearch.4.L012018},
  url = {https://link.aps.org/doi/10.1103/PhysRevResearch.4.L012018}
}

@article{drude_zur_1900,
	title = {On the electron theory of metals, },
	volume = {306},
	issn = {00033804},
	url = {http://doi.wiley.com/10.1002/andp.19003060312},
	doi = {10.1002/andp.19003060312},
	number = {3},
	urldate = {2021-06-07},
	journal = {Annalen der Physik},
	author = {Drude, P.},
	year = {1900},
	pages = {566--613}
}

@article{ciftja_detailed_2020,
	title = {Detailed solution of the problem of {Landau} states in a symmetric gauge},
	volume = {41},
	issn = {0143-0807, 1361-6404},
	url = {https://iopscience.iop.org/article/10.1088/1361-6404/ab78a7},
	doi = {10.1088/1361-6404/ab78a7},
	language = {en},
	number = {3},
	urldate = {2025-01-13},
	journal = {European Journal of Physics},
	author = {Ciftja, Orion},
	month = may,
	year = {2020},
	pages = {035404},
}

@article{kim_terahertz_2023,
	address = {Montreal, QC, Canada},
	title = {Terahertz {Landau} {Polaritons} in {Nano}-slots: {Ultrastrong} {Coupling} under {Extreme} {Spatial} {Confinement}},
	isbn = {979-8-3503-3660-3},
	shorttitle = {Terahertz {Landau} {Polaritons} in {Nano}-slots},
	url = {https://ieeexplore.ieee.org/document/10298888/},
	doi = {10.1109/IRMMW-THz57677.2023.10298888},
	language = {en},
	urldate = {2024-02-19},
	journal = {48th {International} {Conference} on {Infrared}, {Millimeter}, and {Terahertz} {Waves} ({IRMMW}-{THz})},
	publisher = {IEEE},
	author = {Kim, Dasom and Kim, Sunghwan and Lee, Dukhyung and Liang, Shuang and Tay, Fuyang and Manfra, Michael J. and Kim, Dai-Sik and Kono, Junichiro},
	month = sep,
	year = {2023},
	pages = {1--2},
}

@article{krummheuer_pure_2005,
	title = {Pure dephasing and phonon dynamics in {GaAs}- and {GaN}-based quantum dot structures: {Interplay} between material parameters and geometry},
	volume = {71},
	shorttitle = {Pure dephasing and phonon dynamics in {GaAs}- and {GaN}-based quantum dot structures},
	url = {https://link.aps.org/doi/10.1103/PhysRevB.71.235329},
	doi = {10.1103/PhysRevB.71.235329},
	number = {23},
	urldate = {2024-02-07},
	journal = {Physical Review B},
	author = {Krummheuer, B. and Axt, V. M. and Kuhn, T. and D’Amico, I. and Rossi, F.},
	month = jun,
	year = {2005},
	pages = {235329},
}

@book{malic_graphene_2013,
	title = {Graphene and {Carbon} {Nanotubes}},
	language = {en},
	publisher = {John Wiley \& Sons},
    url = {https://www.wiley-vch.de/de?option=com_eshop&view=product&isbn=9783527411610&title=Graphene%20and%20Carbon%20Nanotubes},
	author = {Malic, Ermin and Knorr, Andreas},
	year = {2013},
}

@article{haldane_nobel_2017,
	title = {Nobel {Lecture}: {Topological} quantum matter},
	volume = {89},
	shorttitle = {Nobel {Lecture}},
	url = {https://link.aps.org/doi/10.1103/RevModPhys.89.040502},
	doi = {10.1103/RevModPhys.89.040502},
	number = {4},
	urldate = {2024-10-29},
	journal = {Reviews of Modern Physics},
	author = {Haldane, F. Duncan M.},
	month = oct,
	year = {2017},
	pages = {040502},
}

@article{sondhi_skyrmions_1993,
	title = {Skyrmions and the crossover from the integer to fractional quantum {Hall} effect at small {Zeeman} energies},
	volume = {47},
	url = {https://link.aps.org/doi/10.1103/PhysRevB.47.16419},
	doi = {10.1103/PhysRevB.47.16419},
	number = {24},
	urldate = {2025-07-18},
	journal = {Physical Review B},
	author = {Sondhi, S. L. and Karlhede, A. and Kivelson, S. A. and Rezayi, E. H.},
	month = jun,
	year = {1993},
	pages = {16419--16426},
}

@article{pan_experimental_2008,
	title = {Experimental studies of the fractional quantum {Hall} effect in the first excited {Landau} level},
	volume = {77},
	url = {https://link.aps.org/doi/10.1103/PhysRevB.77.075307},
	doi = {10.1103/PhysRevB.77.075307},
	number = {7},
	urldate = {2024-10-29},
	journal = {Physical Review B},
	author = {Pan, W. and Xia, J. S. and Stormer, H. L. and Tsui, D. C. and Vicente, C. and Adams, E. D. and Sullivan, N. S. and Pfeiffer, L. N. and Baldwin, K. W. and West, K. W.},
	month = feb,
	year = {2008},
	pages = {075307},
}

@article{balram_fractionally_2015,
	title = {Fractionally charged skyrmions in fractional quantum {Hall} effect},
	volume = {6},
	copyright = {2015 The Author(s)},
	issn = {2041-1723},
	url = {https://www.nature.com/articles/ncomms9981},
	doi = {10.1038/ncomms9981},
	language = {en},
	number = {1},
	urldate = {2025-07-18},
	journal = {Nature Communications},
	author = {Balram, Ajit C. and Wurstbauer, U. and Wójs, A. and Pinczuk, A. and Jain, J. K.},
	month = nov,
	year = {2015},
	pages = {8981},
}

@article{papic_fractional_2022,
	title = {Fractional quantum {Hall} effect in semiconductor systems},
	url = {http://arxiv.org/abs/2205.03421},
	doi = {10.48550/arXiv.2205.03421},
	urldate = {2024-10-29},
	journal = {arXiv},
	author = {Papić, Zlatko and Balram, Ajit C.},
	month = may,
	year = {2022},
    pages={arXiv:2205.03421}
}

@article{murthy_hamiltonian_2003,
	title = {Hamiltonian theories of the fractional quantum {Hall} effect},
	volume = {75},
	url = {https://link.aps.org/doi/10.1103/RevModPhys.75.1101},
	doi = {10.1103/RevModPhys.75.1101},
	number = {4},
	urldate = {2024-10-29},
	journal = {Reviews of Modern Physics},
	author = {Murthy, Ganpathy and Shankar, R.},
	month = oct,
	year = {2003},
	pages = {1101--1158},
}

@book{prange_quantum_1990,
	address = {New York, NY},
	series = {Graduate {Texts} in {Contemporary} {Physics}},
	title = {The {Quantum} {Hall} {Effect}},
	copyright = {http://www.springer.com/tdm},
	isbn = {978-0-387-97177-3},
	url = {http://link.springer.com/10.1007/978-1-4612-3350-3},
	urldate = {2025-07-17},
	publisher = {Springer},
	editor = {Prange, Richard E. and Girvin, Steven M. and Birman, Joseph L. and Faissner, H. and Lynn, Jeffrey W.},
	year = {1990},
	doi = {10.1007/978-1-4612-3350-3},
}

@book{girvin_modern_2019,
  title={Modern condensed matter physics},
  author={Girvin, Steven M and Yang, Kun},
  year={2019},
  publisher={Cambridge University Press},
 url = {https://www.cambridge.org/highereducation/books/modern-condensed-matter-physics/F0A27AC5DEA8A40EA6EA5D727ED8B14E#contents},
doi = {10.1017/9781316480649}
}

@article{dung_three-dimensional_1998,
	title = {Three-dimensional quantization of the electromagnetic field in dispersive and absorbing inhomogeneous dielectrics},
	volume = {57},
	issn = {1050-2947, 1094-1622},
	url = {https://link.aps.org/doi/10.1103/PhysRevA.57.3931},
	doi = {10.1103/PhysRevA.57.3931},
	language = {en},
	number = {5},
	urldate = {2021-06-18},
	journal = {Physical Review A},
	author = {Dung, Ho Trung and Knöll, Ludwig and Welsch, Dirk-Gunnar},
	month = may,
	year = {1998},
	pages = {3931--3942},
}

\clearpage

\appendix

\begin{widetext}

\begin{center}
{\bf END MATTER}
\end{center}

\section{Nonlocal susceptibility \label{app: nonlocal susceptibility}}
The longitudinal ($L$), transverse ($T$) and Hall ($H$) entries of the LL susceptibility are derived in the SM \cite{supp} and follow as:
\begin{align}
\chi^L({q},\omega)
    &=\chi_d(\omega)
    +
    \frac{\hbar^2}{\pi (2l_B)^4}\,
\frac{\chi_{\mathrm d}(\omega)}{\bar{n}_{\mathrm{el}} M}
\sum_{\ell,\ell^\prime}
W_{\ell,\ell^\prime}(\omega)
\left[
    \mathcal{Q}_{\ell,\ell^\prime}(q)
    +
\mathcal{Q}_{\ell^\prime,\ell}(q)
\right]^2,\\
\chi^T({q},\omega)
&=
    \chi_d(\omega)
    +
\frac{\hbar^2}{\pi (2l_B)^4}\,
\frac{\chi_{\mathrm d}(\omega)}{\bar{n}_{\mathrm{el}} M}
\sum_{\ell,\ell^\prime}
W_{\ell,\ell^\prime}(\omega)
\left[
\mathcal{Q}_{\ell,\ell^\prime}(q)
-
\mathcal{Q}_{\ell^\prime,\ell}(q)
\right]^2,
\\
\chi^H({q},\omega)
&=
\frac{\hbar^2}{\pi (2l_B)^4}\,
\frac{\chi_{\mathrm d}(\omega)}{\bar{n}_{\mathrm{el}} M}
\sum_{\ell,\ell^\prime}
W_{\ell,\ell^\prime}(\omega)
\left(
\left[\mathcal{Q}_{\ell,\ell^\prime}(q)\right]^2
- 
\left[\mathcal{Q}_{\ell^\prime,\ell}(q)\right]^2
\right)\label{eq: explicit components chi in TE TM basis},
\end{align}
where frequency dependency is given by the Drude susceptibility $\chi_d$ and a Lorentzian LL term $W_{\ell,\ell^\prime}$,
\begin{align}
    \chi_\text{d}(\omega)
    =&\,
    -\frac{\omega_{\text{p}}^2}{\omega(\omega+i\gamma)},
    \qquad
    W_{\ell,\ell^\prime}(\omega)
    =
\frac{f_{\ell}-f_{\ell^\prime}}
{\hbar(\omega+\omega_{\ell}-\omega_{\ell^\prime}+i\gamma)},
\end{align}
and $f_{\ell}$ denotes the temperature-dependent Fermi-Dirac distribution of LL $\ell$.
The wavevector dependency is captured by
\begin{align}
    \mathcal{Q}_{\ell,\ell^\prime}(q)
    &=
\sqrt{\ell+1}
\tilde{F}_{\ell+1, \ell^\prime}\!\left(\frac{l_B}{\sqrt2}  q\right)
+
\sqrt{\ell^\prime}
\tilde{F}_{\ell, \ell^\prime-1}\!\left(\frac{l_B}{\sqrt2}  q\right).
\end{align}
The form factors are \cite{richards_inelastic_2000,oji_magnetoplasma_1986,macdonald_magnetoplasmon_1985,kallin_excitations_1984}
\begin{align}
    \tilde{F}_{\ell , \ell^\prime}\!\left(\frac{l_B}{\sqrt2}  q\right)
    &=
    e^{- q^2 \frac{l_B^2}{4}}
e^{-
\frac12\bigl|\ln\Gamma(\ell+1)-\ln\Gamma(\ell'+1)\bigr|}
\left(\frac{l_B}{\sqrt2} q\right)^{|\ell-\ell'|}
\,
L_{\min(\ell,\ell')}^{|\ell-\ell'|}
\!\left(\frac{l_B^2 q^2}{2}\right),
\end{align}
using the natural logarithm of $\Gamma$-function instead of the formally equivalent factorials to increase numerical stability.
Note that artificially appearing nonphysical form factors $\tilde{F}_{\ell<0, \ell^\prime} = \tilde{F}_{\ell, \ell^\prime<0} = 0$ are set to $0$.
While 
the Fermi–Dirac distribution
for the equilibrium occupations $f_\ell(B)$ is defined from
\begin{align}\label{eq: LL Fermi-Dirac}
    f_\ell(B) = \left(e^{\frac{\hbar \omega_\ell - \mu(T, B)}{k_B T}} + 1\right)^{-1}
    \quad\text{with}\quad
    \mu(T, B)
    = - k_B T \ln\left(\frac{w \cosh\alpha + \sqrt{1 + w^2 \sinh^2 \alpha}}{1 - w}\right)
    + \hbar\omega_c(\ell_F + 1),
\end{align}
where the abbreviations $\alpha = \frac{\hbar\omega_c}{2 k_B T}$ and $w = \ell_F - \nu + 1$.
For low temperatures ($k_B T \ll \hbar\omega_c)$, the chemical potential $\mu(T, B)$ of a 2DEG is derived assuming electron number conservation \cite{vagner_ideally_1983} (see also \cite{meyer_proposal_2026}). The highest (partially) occupied Landau level $\ell_F$ is obtained as the integer part (floor) of the filling factor $\nu =  \bar{n}_\text{el} 2\pi l_B^2$ \cite{jain_composite_2015}.

We obtain the local LL limit for $q=0$. Transformed into the circular basis, where the local LL susceptibility is diagonal (entries: $++$ cyclotron-active; $--$ cyclotron-inactive), this limit gives the common 
{\it local} expression~\cite{wang_direct_2010,jasper_broadband_2020,zhang_superradiant_2014},
\begin{align}
\chi^{\pm\pm}(\mathbf q = \mathbf{0},\omega)
&=
\frac{-\omega_{\text{p}}^2}{\omega\left( \omega +i{\gamma}
    \mp \omega_c  \right)}
, \qquad
\chi^{\pm\mp}(\mathbf q = \mathbf{0},\omega)
=
0.
\end{align}
The full derivation and the real space representation of the nonlocal LL susceptibility are provided in the SM \cite{supp}.

\section{Green's function \label{app: Greens dyad}}
In general, the total (electric-field) Green's function comprises a homogeneous $(0)$ and a scattered $(\text{sc})$ contribution,
\begin{align}
    \mathbcal{G}(\mathbf{r},\mathbf{r}' ;\omega)
    &=
    \mathbcal{G}^{(0)}(\mathbf{r}-\mathbf{r}' ;\omega)
    +
    \mathbcal{G}^{\text{sc}}(\boldsymbol{\rho}-\boldsymbol{\rho}^\prime,z,z' ;\omega),
\end{align}
with the 3D wavevector $k = \sqrt{\varepsilon}\frac{\omega}{c_0}$.
The Green's function for the homogeneous dielectric environment is \cite{jackson_klassische_2014}
\begin{align}
\mathbcal{G}^{(0)}(\mathbf{r} ;\omega) =& \frac{e^{ik|\mathbf{r}|}}{4\pi\varepsilon}
\bigg[
\left(k^2\frac{1}{|\mathbf{r}|} + ik\frac{1}{|\mathbf{r}|^2} - \frac{1}{|\mathbf{r}|^3} \right)\mathbb{1}  \label{eq: dyadic Greens function, real space} 
+ \left(-k^2 - ik\frac{3}{|\mathbf{r}|} + \frac{3}{|\mathbf{r}|^2} \right)
\frac{\mathbf{r}\otimes\mathbf{r} }{|\mathbf{r}|^3}
\bigg].
\end{align}
The retarded scattered Green's function, including the nonlocal LL response is given by
\begin{align*}
    &\mathbcal{G}^{\mathrm{sc}}(\boldsymbol{\rho},z,z';\omega)
    =
\frac{-1}{4\pi\varepsilon}
\int_0^\infty\!\!\!\!\mathrm{d}q\, q\,
k^2
\frac{e^{ik_q\left(|z'|+|z|\right)}}{4\varepsilon \, k_q D_H(q;\omega)}\,
    \begin{pmatrix}
J_0(q\rho)
&
-e^{-2i\phi_{\rho}} J_2(q\rho)
&
-\sqrt{2}\dfrac{q}{k_q}s_{z'}e^{-i\phi_\rho} iJ_1(q\rho)
\\[0.9em]
-e^{2i\phi_{\rho}}  J_2(q\rho)
&
J_0(q\rho)
&
-\sqrt{2}\dfrac{q}{k_q}s_{z'}e^{i\phi_\rho} iJ_1(q\rho)
\\[0.9em]
\sqrt{2}\dfrac{q}{k_q}s_z e^{i\phi_\rho} iJ_1(q\rho)
&
\sqrt{2}\dfrac{q}{k_q}s_z e^{-i\phi_\rho} iJ_1(q\rho)
&
-2\dfrac{q^2}{k_q^2}s_zs_{z'} J_0(q\rho)
\end{pmatrix}
\\[0.5em]
 \circ&
\left[
{ k_q^2 }
\left(
\frac{k_q}{k^2}\chi^L(q;\omega)-\dfrac{i}{2\varepsilon}
\left|\boldsymbol{\chi}(q;\omega)\right|
\right)
\begin{pmatrix}
1 & 1 & 1 \\
1 & 1 & 1 \\
1 & 1 & 1
\end{pmatrix}\right.
\mkern-6mu
{+}
{k^2}
\left(
\frac{\chi^T(q;\omega)}{k_q}-\dfrac{i}{2\varepsilon}
\left|\boldsymbol{\chi}(q;\omega)\right|
\right)
\begin{pmatrix}
1 & -1 & 0 \\
-1 & 1 & 0 \\
0 & 0 & 0
\end{pmatrix}
\mkern-4mu
{+}
\left.
\mkern-2mu
k_q \chi^H(q;\omega)
\begin{pmatrix}
2 & 0 & 1 \\
0 & -2 & -1\\
1 & -1 & 0
\end{pmatrix}
\right],
\nonumber\\[0.5em]
&\qquad\qquad
\end{align*}
where $\circ$ denotes the Hadamard product, i.e., elementwise multiplication, introduced for notational clarity, $|\boldsymbol{\chi}(q,\omega)| = \chi^L(q,\omega)\chi^T(q,\omega)+\left(i\chi^H(q,\omega)\right)^2$
the susceptibility determinant and
$k_q = \sqrt{\frac{\varepsilon}{c_0^2}\omega^2 - {q}^2}$ the out-of-plane wavevector. We further introduced the abbreviations  $s_z:=-\operatorname{sgn}(z)$, $s_{z'}:=-\operatorname{sgn}(z')$ and
\begin{align}
k_q D_H 
=
\left(
1-\frac{i k_q}{2\varepsilon}\chi^L
\right)
\left(
k_q-\frac{i k^2}{2\varepsilon}\chi^T
\right)
-
k_q
\frac{k^2}{4\varepsilon^2}
\left(i\chi^H\right)^2.
\end{align}
The complete derivation is detailed in the SM \cite{supp}.

\end{widetext}

\section{Parameters \label{app sec: parameters}}

As a representative dipole value in the THz regime, we adopt $d=100\, e$nm as the appropriate order of magnitude for THz emitter dipole moments such as graphene quantum dots, which have recently been measured~\cite{messelot_large_2022}.
We also stress that our results for the Purcell factor as well as the qualitative sign change of the Lamb shift are independent on the precise value of the dipole moment.
\vspace{-0.5cm}

\begin{table}[b]
\centering
\caption{Constants and Parameters}
\label{tab:params}
\renewcommand{\arraystretch}{1.2}
\begin{tabular}{lllll}
\hline\hline
Description & Name & Value & Unit & Ref. \\
\hline
Temperature & $T$ & $10$ & $\text{K}$ & \\
Effective mass (GaAs) & $M$ & $0.067\,m_e$ & $\text{fs}^2\,\text{eV}\,\text{nm}^{-2}$ & \cite{krummheuer_pure_2005} \\
2D Electron density& $\bar{n}_{\text{el}}$ & $0.0036$ & $\text{nm}^{-2}$ & \cite{kim_terahertz_2023} \\
Damping & $\hbar\gamma$  & $0.066$  & meV &\cite{curtis_cyclotron_2016,wang_direct_2010}\\
Permittivity (GaAs) & $\varepsilon$ & $ 12.9 $ &  & \\
Dipole moment & $d$ & $100$ & e$\,$nm & \cite{messelot_large_2022}\\
\hline\hline
Speed of light & $c_0$ & $299.7925$ & nm\,fs\(^{-1}\)& \\
Vacuum permittivity & \( \varepsilon_0 \) & $0.05526308$ & e\(^2\)\,eV\(^{-1}\)\,nm\(^{-1}\)  & \\
Planck constant & \( \hbar \) & $0.658212196$ & eV\,fs \\
Boltzmann constant & \( k_B \) & $0.0861745$ & meV\,K\(^{-1}\) & \\
Electron mass & \( m_e \) & $5.6856800$  & fs$^2$ nm$^{-2}$ eV & \\
\hline\hline
Cyclotron frequency & & \multicolumn{3}{l}{$\omega_c = eB/M$} \\
Magnetic length     & & \multicolumn{3}{l}{$l_B = \sqrt{\hbar/(eB)}$} \\
2D plasma frequency & & \multicolumn{3}{l}{$\omega_{\text{p}} = \sqrt{\bar{n}_{\text{el}}\, e^2/(\varepsilon_0 M)}$} \\
Filling factor & & \multicolumn{3}{l}{$\nu =  \bar{n}_\text{el} 2\pi l_B^2$}\\
Highest occupied LL & & \multicolumn{3}{l}{$\ell_F =$ floor$(\nu)$}\\
Larmor radius & & \multicolumn{3}{l}{$r_c = l_B\sqrt{2\ell +1}$}\\
\hline\hline
\end{tabular}
\renewcommand{\arraystretch}{1.0}
\end{table}

\end{document}